\documentclass[12pt,12pt]{article}
\usepackage{graphicx}
\usepackage{epsfig}
\usepackage{amsmath,amssymb}
\usepackage{slashed} 
\catcode`\@=11
\newdimen\ex@
\font\dozeb=cmmib10 scaled \magstep1
\font\dozesyb=cmbsy10 scaled \magstep1
\font\dezb=cmmib10
\def\bm{\fam9}
\def\beq{\begin{equation}}
\def\eeq{\end{equation}}
\def\beqa{\begin{eqnarray}}
\def\eeqa{\end{eqnarray}}
\newcommand\BA{\begin{array}}
\newcommand\EA{\end{array}}

\def\E{{\bm E}}

\begin{document}
\def\thefootnote{\fnsymbol{footnote}}

\title{\bf Extended Supersymmetric $\sigma$-Model\\ 
Based on the SO(2N+1) Lie Algebra\\
of the Fermion Operators\footnotemark[1]}
\vskip1cm
\author
{Seiya NISHIYAMA\footnotemark[2] $\!\!$, 
Jo\~ao da PROVID\^{E}NCIA\footnotemark[3]\\
\\
Constan\c{c}a PROVID\^{E}NCIA\footnotemark[4]~~and 
Fl\' avio CORDEIRO\footnotemark[5]\\
\\
Centro de F\'\i sica Te\' orica,
Departamento de F\'\i sica,\\ 
Universidade de Coimbra\\
P-3004-516 Coimbra, Portugal\\ \\}

\maketitle


\footnotetext[1]
{A preliminary version of
this work has been presented by S. Nishiyama
at the YITP workshop YITP-\\ 
~~~~~~W-07-05 on {\em String Theory and Quantum Field Theory} held at
Kinki University, 6 - 10 August, 2007.}
\footnotetext[2]
{Corresponding author.
~E-mail address: seikoaquarius@ybb.ne.jp,~
nisiyama@teor.fis.uc.pt}
\footnotetext[3]
{E-mail address: providencia@teor.fis.uc.pt}  
\footnotetext[4]
{E-mail address: cp@teor.fis.uc.pt}
\footnotetext[5]
{E-mail address: flaviocordeiro\_704@hotmail.com} 

\begin{abstract}
Extended supersymmetric $\sigma$-model is given, 
standing on 
the $SO(2N \!+\! 1)$ Lie algebra of fermion operators composed of 
annihilation-creation operators and pair operators.
Canonical transformation,
the extension of the $SO(2N)$ Bogoliubov transformation 
to the $SO(2N \!+\! 1)$ group, 
is introduced.
Embedding the $SO(2N \!+\!1)$ group into an $SO(2N \!+\!2)$ group and 
using $\frac{SO(2N \!+\! 2)}{U(N \!+\! 1)}$ coset variables, 
we investigate a new aspect of the supersymmetric $\sigma$-model on
the K\"{a}hler manifold of the symmetric space 
$\frac{SO(2N \!+\! 2)}{U(N \!+\! 1)}$.
We construct a Killing potential which is just the extension of 
the Killing potential in the $\frac{SO(2N)}{U(N)}$ coset space 
given by van Holten et al.
to that in the $\frac{SO(2N \!+\! 2)}{U(N \!+\! 1)}$ coset space.
To our great surprise,
the Killing potential is equivalent with the generalized density matrix. 
Its diagonal-block matrix is related to a reduced scalar potential 
with a Fayet-Ilipoulos term.
The reduced scalar potential is optimized 
in order to see the behaviour of
the vacuum expectation value of the $\sigma$-model fields
and a proper solution for one of the $SO(2N \!+\! 1)$ group parameters
is obtained.
We give bosonization  
of the $SO(2N \!+\! 2)$ Lie operators,
vacuum functions and differential forms for their bosons expressed
in terms of the $\frac{SO(2N+2)}{U(N+1)}$ coset variables,
a $U(1)$ phase
and the corresponding K\"{a}hler potential.
\end{abstract}


\newpage

\setcounter{equation}{0}
\renewcommand{\theequation}{\arabic{section}.\arabic{equation}}

\section{Introduction}

~~~~
The supersymmetric extension of nonlinear models 
was first given by Zumino under the introduction of scalar fields
taking values in a complex K\"{a}hler manifold
\cite{Zumino.79}.
The higher dimensional nonlinear $\sigma$-models 
defined on symmetric spaces and on hyper K\"{a}hler manifolds
have been itensively studied in various contexts 
in modern versions of elementary particle physics,
superstring theory and supergravity theory
\cite{NNH.01,HKN.02,LRHKLR.8387,ANNS.03}.

In nuclear and condensed matter physics,
the time dependent Hartree-Bogoliubov (TDHB) theory
\cite{BCS.57,Bog.59}
has been regarded as the standard approximation 
in the many-body theoretical descriptions of 
superconducting fermion systems
\cite{RS.80,BR.86}.
In the TDHB
an HB wavefunction (WF) for such systems represents 
Bose condensate states of fermion pairs. 
It is a good approximation for the ground state of the system 
with a short-range pairing interaction
that produces a spontaneous Bose condensation of the fermion pairs. 
Number-nonconservation of the HB WF is 
a consequence of the spontaneous Bose condensation of fermion pairs.
Standing on the Lie-algebraic viewpoint,
the  pair operators of fermion with $N$ indices 
form the $SO(2N)$ Lie algebra
and accompany the $U(N)$ Lie algebra as a sub-algebra. 
$SO(2N)( \!=\! g)$ and $U(N)( \!=\! h)$ denote the special orthogonal group 
of $2N$ dimensions and the unitary group of $N$ dimensions, respectively.
One can give an integral representation of a state vector on the group $g$,
the exact coherent state representation (CS rep) of a fermion system
\cite{Perelomov.86}.
It makes possible global approach to the above problem. 
The canonical transformation of the fermion operators generated by 
the Lie operators in the $SO(2N)$ Lie algebra induces 
the well-known generalized Bogoliubov transformation for the fermions. 
The TDHB equation has been derived 
from the Euler-Lagrange equation of motion (EOM)
for the $\frac{g}{h} \!=\! \frac{SO(2N)}{U(N)}$ coset variables
by one of the present authors (SN)
\cite{Ni.81.82}.
The TDHB theory is, however, applicable only to even fermion systems. 
For odd fermion systems 
we have no TD self-consistent field (SCF) theory
with the same level of the mean field (MF) approximation 
as the TDHB. 

van Holten et al.
have discussed a procedure for consistent coupling of gauge- and 
matter superfields to supersymmetric $\sigma$-models 
on the K\"{a}hler coset spaces.
They have presented a way of constructing 
the Killing potentials and have applied their method
to the explcit construction of supersymmetric $\sigma$-models 
on the coset spaces $\frac{SO(2N)}{U(N)}$.
They have shown that 
only a finite number of the coset models can be consistent 
when coupled to matter superfields with $U(N)$ quantum numbers
reflecting spinorial representations of $SO(2N)$
\cite{NNH.01}.
Higashijima et al. 
have given Ricci-flat metrics on 
compact K\"{a}hler manifolds, 
$\frac{SU(N)}{[SU(N-M) \times U(M)]}$, $\frac{SO(2N)}{U(N)}$ and 
$\frac{Sp(N)}{U(N)}$
and non-compact K\"{a}hler manifolds,
applying their technique of the gauge theory formulation of 
supersymmetric nonlinear $\sigma$-models on the hermitian symmetric spaces
\cite{HKN.02}.
Preceding these works,
Deldug and Valent have investigated the K\"{a}hlerian $\sigma$-models
in two-dimensional space-time at the classical quantum level.
They have presented a unified treatment of the models
based on irreducible hermitian symmetric spaces 
corresponding to the coset spaces $\frac{G}{H}$
\cite{DelVal.85}.
van Holten has also discussed the construction of $\sigma$-models
on compact and non-compact Grassmannian manifolds,
$\frac{SU(N+M)}{S[U(N) \times U(M)]}$ and 
$\frac{SU(N,M)}{S[U(N) \times U(M)]}$
\cite{vanHolten.85}.

One of the most challenging problems in the current studies of nuclear
physics is to give a theory suitable for description of collective
motions with large amplitudes in both even and odd soft nuclei with
strong collective correlations. 
For providing the general microscopic means 
for a unified self-consistent description 
for Bose and Fermi type collective excitations in such fermion systems,  
Fukutome, Yamamura and one of the present authors (SN) have proposed
a new fermion many-body theory
basing on the $SO(2N \!+\! 1)$ Lie algebra of fermion operators 
\cite{FYN.77}.
The set of the fermion operators 
composed of creation-annihilation and pair operators forms 
a larger Lie algebra, 
the Lie algebra of the $SO(2N \!+\! 1)$ group.
A representation of an $SO(2N \!+\! 1)$ group has been derived 
by a group extension of the $SO(2N)$ Bogoliubov transformation 
for fermions 
to a new canonical transformation group.  
The fermion Lie operators, when operated onto 
the integral representation of
the $SO(2N \!+\! 1)$ WF, are mapped into
the regular representation of the $SO(2N \!+\! 1)$ group and 
are represented by boson operators. 
The boson images of the fermion Lie operators
are expressed by closed first order differential forms. 
The creation-annihilation operators themselves 
as well as the pair operators 
are given by the Schwinger-type boson representation 
as a natural consequence
\cite{Sch.65,YN.76}.

Along the same way as the above,
we give an extended supersymmetric $\sigma$-model
on K\"{a}hler coset space 
$\!\frac{G}{H} \!=\!\! \frac{SO(2N \!+\! 2)}{U(N \!+\! 1)}\!$,
basing on the $SO(2N \!+\! 1)$ Lie algebra of the fermion operators.
Embedding the $SO(2N \!+\! 1)$ group into an $SO(2N \!+\! 2)$ group and 
using the $\!\frac{SO(2N \!+\! 2)}{U(N \!+\! 1)}\!$ coset variables  
\cite{Fuk.77},
we investigate a new aspect of the supersymmetric $\sigma$-model on
the K\"{a}hler manifold of the symmetric space 
$\!\frac{SO(2N \!+\! 2)}{U(N \!+\! 1)}\!$.
We construct a Killing potential which is just the extension of 
the Killing potential in the $\!\!\frac{SO(2N)}{U(N)}\!\!$ coset space 
given by van Holten et al. $\!\!\!$
\cite{NNH.01} \hspace{-0.3cm} 
to that in the $\!\!\frac{SO(2N \!+\! 2)}{U(N \!+\! 1)}\!\!$ coset space.
To our great surprise,
the Killing potential is equivalent with the generalized density matrix. 
Its diagonal-block part is related to a reduced scalar potential 
with a Fayet-Ilipoulos term.
The reduced scalar potential is optimized 
in order to see the behaviour of
the vacuum expectation value of the $\sigma$-model fields.
We get, however, a too simple solution.
The optimiztion of the reduced scalar potential
plays an important role to evaluate
the criteria for supersymmetry-breaking and internal symmetry-breaking.
To find a proper solution for the extended supersymmetryic $\sigma$-mdoel,
after rescaling Goldstone fields by a mass parameter, 
minimization of the deformed reduced scalar potential is also made.
 
Previously,
using the above embedding 
we have developed 
an extended TDHB (ETDHB) theory,
in which paired and unpaired modes are treated on the same footing 
\cite{FYN.77,Ni.81.82}. 
The ETDHB applicable to both even and odd fermion systems 
is a TDSCF 
with the same level of MF approximation 
as the usual TDHB for even fermion systems
\cite{BR.86}. 
The ETDHB equation is derived from 
a classical Euler-Lagrange EOM 
for the $\frac{SO(2N \!+\! 2)}{U(N \!+\! 1)}$ coset variables. 

In sect. 2,
we recapitulate briefly an induced representation 
of an $SO(2N \!+\! 1)$ canonical transformation group.
In sect. 3,
an embedding of the $SO(2N \!+\! 1)$ group into an $SO(2N \!+\! 2)$ group 
is made 
and introduction of 
$\frac{SO(2N \!+\! 2)}{U(N \!+\! 1)}$ coset variables is also made.
In sect. 4,
we give an extended supersymmetric $\sigma$-model
on the coset space $\frac{SO(2N \!+\! 2)}{U(N \!+\! 1)}$
based on the $SO(2N \!+\! 1)$ Lie algebra of the fermion operators
and study a new aspect of the extended supersymmetric $\sigma$-model on the K\"{a}hler manifold, 
the symmetric space $\frac{SO(2N \!+\! 2)}{U(N \!+\! 1)}$.
In sect. 5,
the expressions for Killing potentials
in 
that coset space are given and 
their equivalence with the generalized density matrix is proved
and then a reduced scalar potential is derived.
Finally, in sect. 6, 
we give 
discussions on the optimized scalar potential
and concluding remarks. 
In Appendix,
we give
a bosonization 
of the $SO(2N \!+\! 2)$ Lie operators,
vacuum functions and differential forms for their bosons expressed 
in terms of the corresponding K\"{a}hler potential, 
the $\frac{SO(2N+2)}{U(N+1)}$ coset variables
and a $U(1)$ phase
and 
make
a brief sketch of derivation of the ETDHB equation 
from the Euler-Lagrange EOM 
for 
those coset variables in the TDSCF.
Throughout this paper, we use the summation convention over
repeated indices unless there is the danger of misunderstanding.


\newpage

\setcounter{equation}{0}
\renewcommand{\theequation}{\arabic{section}.\arabic{equation}}

\section{The SO(2N+1) Lie algebra of fermion operators and the Bogoliubov transformation}

\def\bra#1{{<\!#1\,|}} 
\def\ket#1{{|\,#1\!>}}

~~~~Let $c_{\alpha }$ and $c^{\dag }_{\alpha }$, $\alpha$ 
\!=\! 
1,$\cdot \cdot \cdot$, 
$N$, be annihilation and creation operators of the fermion
satisfying the canonical anti-commutation relations
\beq
\{c_{\alpha },~c^{\dag }_{\beta }\}
=
\delta_{\alpha \beta } ,~~
\{c_{\alpha },~c_{\beta }\}
=
\{c^{\dag }_{\alpha },~c^{\dag }_{\beta }\}
=0 .
\label{anticommrel}
\eeq
We introduce the set of fermion operators consisting of the following
annihilation and creation operators and pair operators: 
\beqa
\left.
\BA{ll}
&c_{\alpha },~c^{\dag }_{\alpha } ,\\
\\ 
&E^{\alpha }_{~\beta }
=
c^{\dag }_{\alpha }c_{\beta }
{\displaystyle -\frac{1}{2}} \delta_{\alpha \beta } ,~~
E^{\alpha \beta }
=
c^{\dag }_{\alpha }c^{\dag }_{\beta } ,~~
E_{\alpha \beta }
=
c_{\alpha }c_{\beta } ,\\
\\
&E^{\alpha \dag }_{~\beta }
=
E^{\beta }_{~\alpha } ,~~
E^{\alpha \beta } 
=
E^{\dag }_{\beta \alpha } ,~~
E_{\alpha \beta }
=
- E_{\beta \alpha } .~~
(\alpha , \beta = 1, \cdot \cdot \cdot, N)
\EA
\right\}
\label{operatorset}
\eeqa
It is well known that the set of fermion operators
(\ref{operatorset})
form an $SO(2N+1)$ Lie algebra.
As a consequence of the anti-commutation relation
(\ref{anticommrel}),
the commutation relations for the fermion operators
(\ref{operatorset})
in the $SO(2N+1)$ Lie algebra are 
\beqa
[E^{\alpha }_{~\beta },~E^{\gamma }_{~\delta }]
=
\delta_{\gamma \beta }E^{\alpha }_{~\delta } 
- 
\delta_{\alpha \delta }E^{\gamma }_{~\beta },~~
(U(N)~\mbox{algebra})
\label{commurel1}
\eeqa
\vspace{-0.5cm}
\beqa
\left.
\BA{ll}
&[E^{\alpha }_{~\beta },~E_{\gamma \delta }]
=
\delta_{\alpha \delta }E_{\beta \gamma } 
- 
\delta_{\alpha \gamma }E_{\beta \delta },\\
\\
&[E^{\alpha \beta },~E_{\gamma \delta }]
=
\delta_{\alpha \delta }E^{\beta }_{~\gamma } 
+ 
\delta_{\beta \gamma }E^{\alpha }_{~\delta }
-
\delta_{\alpha \gamma }E^{\beta }_{~\delta } 
- 
\delta_{\beta \delta }E^{\alpha }_{~\gamma },\\
\\
&[E_{\alpha \beta },~E_{\gamma \delta }]
=
0,
\EA
\right\}
\label{commurel2}
\eeqa
\vspace{-0.3cm}
\beqa
\left.
\BA{ll}
&[c^{\dag }_{\alpha },~c_{\beta }]
=
2E^{\alpha }_{~\beta },~
[c_{\alpha },~c_{\beta }]
=
2E_{\alpha \beta },\\
\\
&[c_{\alpha },~E^{\beta }_{~\gamma }]
=
\delta_{\alpha \beta }c_{\gamma },~
[c_{\alpha },~E_{\beta \gamma }]
=
0,\\
\\
&[c_{\alpha },~E^{\beta \gamma }]
=
\delta_{\alpha \beta }c^{\dag }_{\gamma }
-
\delta_{\alpha \gamma }c^{\dag }_{\beta } .
\EA
\right\}
\label{commurel3}
\eeqa
We omit the commutation relations obtained 
by hermitian conjugation of
(\ref{commurel2}) and (\ref{commurel3}).
The $SO(2N+1)$ Lie algebra of the fermion operators contains
the $U(N)(=\{E^{\alpha }_{~\beta }\})$ and 
the $SO(2N)(=\{E^{\alpha }_{~\beta },
~E^{\alpha \beta },~E_{\alpha \beta }\})$ 
Lie algebras of the pair operators 
as sub-algebras.

An $SO(2N)$ canonical transformation $U(g)$ is generated by 
the fermion $SO(2N)$ Lie operators.
The $U(g)$ is the generalized Bogoliubov transformation 
\cite{Bog.59} 
specified by an $SO(2N)$ matrix $g$
\beq
U(g)(c, c^{\dag })U^{\dag }(g)
=
(c, c^{\dag }) g ,
\label{Bogotrans}
\eeq
\beq
g
\stackrel{\mathrm{def}}{=}
\left[ 
\BA{cc} 
a & \bar{b} \\
b & \bar{a} \\ 
\EA 
\right] ,
~~~~
g^{\dag }g = gg^{\dag } = 1_{2N} ,~~~
\det g
=
1 ,
\label{RepBogotrans}
\eeq
\beq
U(g)U(g') = U(gg') ,~~~
U(g^{-1}) = U^{-1}(g) = U^{\dag }(g) ,~~~
U(1_{2N}) = \mathbb{I}_g~(\mbox{unit operator on}~g),
\label{Ug}
\eeq
where ($c$, $c^{\dag }$) is 
the 2$N$-dimensional row vector 
(($c_{\alpha }$), ($c^{\dag }_{\alpha }$)) and 
$a \!=\! (a^{\alpha }_{~\beta })$ and $b \!=\! (b_{\alpha \beta })$ 
are $N \!\times\! N$ matrices. 
The bar denotes the complex conjugation.
The HB ($SO(2N)$) WF $\ket g$ is generated as 
$\ket g \!=\! U(g) \ket 0$ where 
$\ket 0$ is the vacuum satisfying 
$c_{\alpha }\ket 0 \!=\! 0$. 
The matrix $g$ is composed of the matrices $a$ and $b$ satisfying 
the ortho-normalization condition.
The $\ket g$ is expressed as
\beqa 
\ket g
=
\bra 0 U(g) \ket 0
\exp(\frac{1}{2}\cdot q_{\alpha \beta }c^\dagger_\alpha 
c^\dagger_\beta) \ket 0 ,
\label{Bogoketg}
\eeqa
\vskip-1.6\bigskipamount
\beqa
\bra 0 U(g) \ket 0
=
\bar{\Phi }_{00}(g)
=
\left[\det(a)\right]^{\frac{1}{2}}
=
\left[\det(1_N + q^\dag q)\right]^{-\frac{1}{4}}
e^{i\frac{\tau }{2}}~, 
\label{Bogowf}
\eeqa
\vskip-1.6\bigskipamount
\beqa
q = ba^{-1} = -q^{\mbox{\scriptsize T}},~
\mbox{(variables of the $\frac{SO(2N)}{U(N)}$ coset space)} ,~
\tau 
=
\frac{i}{2} \ln \!
\left[\frac{\det({a}^*)}{\det({a})}
\right] ,
\label{Bogocoset}
\eeqa
where $\det$ means determinant and the 
symbol {\scriptsize T} denotes the transposition. 

The canonical anti-commutation relation gives us not only 
the above Lie algebras but also the other three algebras.
Let $n$ be the fermion number operator 
$n \!=\! c^\dag _{\alpha } c_\alpha$.
The operator $(-1)^n$ anticommutes with 
$c_\alpha$ and $c^\dag _\alpha$;
\beq
\{ c_\alpha,~(-1)^n \}
=
\{ c^\dag _\alpha,~(-1)^n \}
=
0.
\label{chiralop}
\eeq
Let us introduce the operator
$
\Theta
$
defined by
$
\Theta
\!\equiv\!
\theta_\alpha c^\dag_\alpha - \bar{\theta }_\alpha c_\alpha 
$.
Due to the relation
$
\Theta ^2
\!=\!
-
\bar{\theta }_\alpha \theta_\alpha
$,
we have
\beqa
\left.
\BA{ll}
e^\Theta
=
Z 
+ X_\alpha c^\dag_\alpha
- \bar{X}_\alpha c_\alpha ,~~
\bar{X}_\alpha X_\alpha + Z^2 = 1 ,\\
\\[-8pt]
Z
=
\cos \theta ,~~
X_\alpha
=
{\displaystyle \frac{\theta_\alpha }{\theta }} \sin \theta ,~~
\theta ^2
=
\bar{\theta }_\alpha \theta_\alpha .
\EA
\right\}
\label{theta}
\eeqa
From
(\ref{anticommrel}), (\ref{chiralop}) and (\ref{theta}),
we obtain
\beqa
\left.
\BA{ll}
&e^\Theta (c_\alpha, c^\dag _\alpha ,
{\displaystyle \frac{1}{\sqrt{2}}}) (-1)^n e^{-\Theta }
=
(c_\beta, c^\dag _\beta ,
{\displaystyle \frac{1}{\sqrt{2}}}) (-1)^n G_X ,\\
\\[-8pt]
&G_X 
\stackrel{\mathrm{def}}{=}
\left[ 
\BA{ccc} 
\delta_{\beta \alpha } 
- 
\bar{X}_\beta X_\alpha &
\bar{X}_\beta \bar{X}_\alpha & -\sqrt{2}Z \bar{X}_\beta \\
\\[-8pt]
X_\beta X_\alpha & \delta_{\beta \alpha } 
- 
X_\beta \bar{X}_\alpha & 
\sqrt{2}ZX_\beta  \\
\\[-8pt]
\sqrt{2}ZX_\alpha & -\sqrt{2}Z \bar{X}_\alpha & 2Z^2 - 1 
\EA 
\right] .
\EA
\right\}
\label{chiraloptrans}
\eeqa
Let $G$ be the $(2N+1) \times (2N+1)$ matrix defined by
\beqa
\!\!\!\!\!\!\!\!
\left.
\BA{ll}
&G
\stackrel{\mathrm{def}}{=}
G_X \!
\left[ \!
\BA{ccc} 
a & \bar{b} & 0 \\
\\[-8pt]
b & \bar{a} & 0 \\
\\[-8pt]
0 & 0 & 1
\EA \!
\right]
\!=\!  
\left[ \!
\BA{ccc} 
a - \bar{X} Y & \bar{b} + \bar{X} \bar{Y} & -\sqrt{2}Z \bar{X} \\
\\[-8pt]
b + X Y & \bar{a} - X \bar{Y} & \sqrt{2}ZX  \\
\\[-8pt]
\sqrt{2}ZY & -\sqrt{2}Z \bar{Y} & 2Z^2 - 1
\EA \!
\right] ,
\BA{c}
X_\alpha
\!=\!
\bar{a}^{\alpha }_{~\beta } Y_\beta 
- 
b_{\alpha \beta } \bar{Y}_\beta ,\\
\\[-8pt]
Y_\alpha
\!=\!
X_\beta a^\beta_{~\alpha } 
- 
\bar{X}_\beta b_{\beta \alpha } ,\\
\\[-8pt]
\bar{Y}_\alpha Y_\alpha + Z^2 = 1 ,
\EA
\EA \!\!
\right\}
\label{defG} 
\eeqa
where $X$ and $Y$ are the column vector and the row vector, respectively.
The $SO(2N+1)$ canonical transformation $U(G)$ is generated by 
the fermion $SO(2N+1)$ Lie operators.
The $U(G)$ is an extension of 
the generalized Bogoliubov transformation $U(g)$ 
\cite{Bog.59} 
to a nonlinear transformation and is specified 
by the $SO(2N+1)$ matrix $G$.
We identify this $G$ with the argument $G$ of $U(G)$.
Then $U(G) \!=\! U(G_X) U(g)$ and $U(G_X) \!=\! \exp (\Theta )$.

From
(\ref{Bogotrans}), (\ref{chiraloptrans}) and (\ref{defG})
and the commutability of $U(g)$ with $(-1)^n$,
we obtain
\beqa
U(G)(c_\alpha, c^{\dag }_\alpha ,\frac{1}{\sqrt{2}}) (-1)^n U^{\dag }(G)
=
(c_\beta, c^{\dag }_\beta, \frac{1}{\sqrt{2}}) (-1)^n
\left[ 
\BA{ccc} 
A_{\beta \alpha } & \bar{B}_{\beta \alpha } & 
{\displaystyle -\frac{\bar{x}_\beta }{\sqrt{2}}} \\
B_{\beta \alpha } & \bar{A}_{\beta \alpha } & 
{\displaystyle \frac{x_\beta }{\sqrt{2}}} \\
{\displaystyle \frac{y_\alpha }{\sqrt{2}}} & 
{\displaystyle -\frac{\bar{y}_\alpha }{\sqrt{2}}} & z 
\EA 
\right] ,
\label{SO2Nplus1chiraltrans}
\eeqa
where
\beqa
\left.
\BA{ll}
&
A_{\alpha \beta }
=
a_{\alpha \beta }
-
\bar{X}_\alpha Y_\beta
=
a_{\alpha \beta }
-
{\displaystyle \frac{\bar{x}_\alpha y_\beta }{2(1 + z)}} ,\\
\\[-8pt]
&
B_{\alpha \beta }
=
b_{\alpha \beta }
+
X_\alpha Y_\beta
=
b_{\alpha \beta }
+
{\displaystyle \frac{x_\alpha y_\beta }{2(1 + z)}} ,\\
\\[-8pt]
&
x_\alpha
=
2ZX_\alpha ,~
y_\alpha
=
2ZY_\alpha ,~
z
=
2Z^2 - 1 .
\EA
\right\}
\label{relAtoaXY}
\eeqa
By using the relation 
$
U(G)(c, c^{\dag },\frac{1}{\sqrt{2}}) U^{\dag }(G)
=
U(G)(c, c^{\dag },\frac{1}{\sqrt{2}}) U^{\dag }(G)
(z + \rho)(-1)^n
$ 
and
the third column equation of
(\ref{SO2Nplus1chiraltrans}),
Eq.
(\ref{SO2Nplus1chiraltrans})
can be written as
\beq
U(G)(c, c^{\dag },\frac{1}{\sqrt{2}}) U^{\dag }(G)
=
(c, c^{\dag }, \frac{1}{\sqrt{2}}) 
(z - \rho)G ,
\eeq
\beq
G 
\stackrel{\mathrm{def}}{=}
\left[ 
\BA{ccc} 
A & \bar{B} & {\displaystyle -\frac{\bar{x}}{\sqrt{2}}} \\
B & \bar{A} & {\displaystyle \frac{x}{\sqrt{2}}} \\
{\displaystyle \frac{y}{\sqrt{2}}} & 
{\displaystyle -\frac{\bar{y}}{\sqrt{2}}} & z 
\EA 
\right],
~~~~
G^{\dag }G = GG^{\dag } = 1_{2N+1} ,~~~
\det G
=
1 ,
\eeq
\beq
U(G)U(G') = U(GG') ,~~~
U(G^{-1}) = U^{-1}(G) = U^{\dag }(G) ,~~~
U(1_{2N+1}) = \mathbb{I}_G ,
\eeq
where ($c$, $c^{\dag }$, $\frac{1}{\sqrt{2}}$) is 
a (2$N$+1)-dimensional row vector 
(($c_{\alpha }$), ($c^{\dag }_{\alpha }$), $\frac{1}{\sqrt{2}}$) and 
$A \!=\! (A^{\alpha }_{~\beta })$ and $B \!=\! (B_{\alpha \beta })$ 
are $N \!\times\! N$ matrices. 
The $U(G)$ is a nonlinear transformation 
with a $q$-number gauge factor $z - \rho$ 
where 
${\rho } 
\!=\! 
x_{\alpha }c^{\dag }_{\alpha }-\bar{x}_{\alpha }c_{\alpha }$ and 
${\rho }^{2} 
\!=\! 
- \bar{x}_{\alpha }x_{\alpha } \!=\! {z}^{2}-1$ 
\cite{FYN.77}.
The matrix $G$ is a matrix belonging to the $SO(2N + 1)$ group.
It can be transformed to a real $(2N + 1)$-dimensional orthogonal matrix
by the transformation
\beq
O = VGV^{-1} ,~~
V
= 
\left[ 
\BA{ccc} 
{\displaystyle \frac{1}{\sqrt{2}}\cdot 1_N}&
{\displaystyle \frac{1}{\sqrt{2}}\cdot 1_N}&0 \\
-{\displaystyle \frac{i}{\sqrt{2}}\cdot 1_N}&
{\displaystyle \frac{i}{\sqrt{2}}\cdot 1_N}&0
\\
0&0&1
\EA
\right] .
\eeq

When $z \!=\! 1$, the $G$ becomes an $SO(2N)$ matrix $g$.  
The $SO(2N+1)$ WF $\ket G \!=\! U(G) \ket 0$ 
is expressed as
\cite{Fuk.77,Fuk.81}
\beqa
\left.
\BA{rl}
\ket G
=&\!\!\!
\bra 0 U(G) \ket 0 (1 + r_\alpha c^\dagger_\alpha)
\exp({\displaystyle \frac{1}{2}} 
\cdot q_{\alpha\beta }c_\alpha^\dagger c_\beta^\dagger) 
\ket0 ,\\
\\[-8pt]
r_\alpha
=&\!\!\!
{\displaystyle \frac{1}{1+z}}
(x_\alpha + q_{\alpha \beta } \bar{x}_\beta) ,
\EA
\right\}
\label{SO2Nplus1wf}
\eeqa
\beqa
\bra0\, U(G)\,\ket0 
= 
\bar{\Phi }_{00}(G) 
= 
\sqrt{\frac{1+z}{2}}
\left[
\det(1_N + q^\dag q)
\right]^{-\frac{1}{4}}
e^{i\frac{\tau }{2}} .
\label{SO2Nplus1vacuumf}
\eeqa


\newpage

\setcounter{equation}{0}
\renewcommand{\theequation}{\arabic{section}.\arabic{equation}}

\section{Embedding into an SO(2N+2) group}

~~~~Following Fukutome
\cite{Fuk.81},
we define the projection operators $P_{\!+}$ and $P_{\!-}$ 
onto the sub-spaces of
even and odd fermion numbers, respectively, by\\[-8pt]
\beq
P_\pm
\stackrel{\mathrm{def}}{=}
{\displaystyle \frac{1}{2}}(1 \pm (-1)^n) ,~~
P_\pm ^2
=
P_\pm ,~~
P_+ P_-
=
0 ,
\eeq
and define the following operators with the supurious index 0:
\beqa
\left.
\BA{ll}
&E^0_{~0}
\stackrel{\mathrm{def}}{=}
-
{\displaystyle \frac{1}{2}} (-1)^n
=
{\displaystyle \frac{1}{2}} (P_- - P_+) ,\\
\\[-6pt]
&E^\alpha_{~0}
\stackrel{\mathrm{def}}{=}
c^\dagger_\alpha P_- 
=
P_+ c^\dagger_\alpha,~~
E^0_{~\alpha }
\stackrel{\mathrm{def}}{=}
c_\alpha P_+ 
=
P_-c_\alpha ,\\
\\[-6pt]
&E^{\alpha 0}
\stackrel{\mathrm{def}}{=}
-c_\alpha^\dagger P_+ 
=
-P_- c_\alpha^\dagger,~~
E^{0 \alpha }
\stackrel{\mathrm{def}}{=}
-E^{\alpha 0} ,\\
\\[-6pt]
&E_{\alpha 0}
\stackrel{\mathrm{def}}{=}
c_\alpha P_- 
=
P_+ c_\alpha ,~~
E_{0 \alpha }
\stackrel{\mathrm{def}}{=}
-E_{\alpha 0} .
\EA
\right\}
\label{spuriousoperators}
\eeqa
The annihilation-creation operators can be expressed 
in terms of the operators
(\ref{spuriousoperators})
as
\beq
c_\alpha
=
E_{\alpha 0} + E^0_{~\alpha },~~
c^\dag_\alpha
=
-E^{\alpha 0} + E^\alpha_{~0} .
\eeq
We introduce the indices $p,~q,~\cdots$ running over  
$N+1$ values $0,1,\cdots,~N$.
Then the operators of 
(\ref{operatorset}) and (\ref{spuriousoperators})
can be denoted in a unified manner as
$E^p_{~q},~E_{pq}$ and $E^{pq}$.
They satisfy
\beqa
\left.
\BA{ll}
&
E^{p \dag }_{~q}
=
E^q_{~p} ,~~
E^{p q } 
=
E^{\dag }_{q p } ,~~
E_{p q }
=
- E_{q p } ,~~
(p , q = 0, 1, \cdots,~N) \\
\\[-6pt]
&
[E^p_{~q},~E^r_{~s}]
=
\delta_{q r}E^p_{~s} 
- 
\delta_{p s}E^r_{~q},~~
(U(N+1)~\mbox{algebra}) \\
\\[-6pt]
&
\left.
\BA{ll}
&[E^p_{~q},~E_{r s}]
=
\delta_{p s}E_{q r} 
- 
\delta_{p r}E_{q s},\\
\\[-6pt]
&[E^{p q},~E_{r s}]
=
\delta_{p s}E^q_{~r}
+ 
\delta_{q r}E^p_{~s}
-
\delta_{p r}E^q_{~s} 
- 
\delta_{q s}E^p_{~r},\\
\\[-6pt]
&[E_{pq},~E_{rs}]
=
0 .
\EA
\right]
\EA
\right\}
\label{commurelp}
\eeqa
The above commutation relations in
(\ref{commurelp}) 
are of the same form as 
(\ref{commurel1}) and (\ref{commurel2}).

Instead of 
(\ref{spuriousoperators}),
it is possible to employ the operators
\beqa
\tilde{E}^0_{~0}
=
{\displaystyle \frac{1}{2}} (-1)^n
=
{\displaystyle \frac{1}{2}} (P_+ - P_-) ,~~
\tilde{E}^\alpha_{~0}
=
c^\dagger_\alpha P_+ ,~~
\tilde{E}^0_{~\alpha }
=
c_\alpha P_- .
\label{spuriousoperators2}
\eeqa
Denoting
$E^\alpha_{~\beta } \equiv \tilde{E}^\alpha_{~\beta }$,
it is shown that the operators 
$\tilde{E}^p_{~q},~p,~q = 0,~1,~\cdots,~N$,
satisfy
\beq
\tilde{E}^{p \dag }_{~q}
=
\tilde{E}^q_{~p},~~
[\tilde{E}^p_{~q},~\tilde{E}^r_{~s}]
=
\delta_{q r}\tilde{E}^p_{~s} 
- 
\delta_{p s}\tilde{E}^r_{~q}.~~
(\tilde{U}(N+1)~\mbox{algebra})
\label{commurel1tilde}
\eeq
The Lie algebra $\tilde{U}(N+1)$ 
is a 
$U(N+1)$ Lie algebra but
it is not unitarily equivalent to 
$U(N+1)$.

Two Clifford algebras $C_{2N}$ and $C_{2N+1}$
can be constructed from the fermion operators:
\beq
M_\alpha
=
c_\alpha + c^\dag_\alpha ,~~
M_{\alpha + N}
=
i(c_\alpha - c^\dag_\alpha ) ,
\eeq
\beq
\bar{M}_\alpha
=
(c_\alpha - c^\dag_\alpha ) (-1)^n,~~
\bar{M}_{\alpha + N}
=
i(c_\alpha + c^\dag_\alpha ) (-1)^n,~~
\bar{M}_0
=
(-1)^n .
\eeq
These operators satisfy
\beq
\{ M_i,~M_j \}
=
2 \delta _{i j} ,~~(i,~j = 1,~\cdots,~2N)
\eeq
\beq
\{ \bar{M}_i,~\bar{M}_j \}
=
2 \delta _{i j} .~~(i,~j = 0,~1,~\cdots,~2N)
\eeq
$C_{2N} = \{ M_i; i = 1,~\cdots,~2N \}$
and
$C_{2N+1} = \{ \bar{M}_i; i = 0,~1,~\cdots,~2N \}$
are the Clifford algebras of $2N$ and $2N+1$ dimensions, respectively
\cite{Murnaghan.38,Chevalley.54}.
They provide the bases to characterize the canonical transformations
generated by the $SO(2N)$ and $SO(2N+1)$ Lie algebras.  

The $SO(2N+1)$ group is embedded into an $SO(2N+2)$ group. 
The embedding leads us to an unified formulation of the $SO(2N+1)$
regular representation in which paired and unpaired modes are
treated in an equal way. 
Define 
($N$+1)$\times$($N$+1) matrices ${\cal A}$ and ${\cal B}$ as  
\beq
{\cal A}
= 
\left[ 
\BA{cc}
A & {\displaystyle -\frac{\bar{x}}{2}} \\
{\displaystyle \frac{y}{2}} & {\displaystyle \frac{1+z}{2}}
\EA 
\right],
~~~
{\cal B}
= 
\left[ 
\BA{cc}
B & {\displaystyle \frac{x}{2}} \\
{\displaystyle -\frac{y}{2}} & {\displaystyle \frac{1-z}{2}}
\EA 
\right],~~~
y = x^{\mbox{\scriptsize T}}a - x^{\dag }b.
\label{calAcalB}
\eeq
Imposing the ortho-normalization of the $G$, 
matrices ${\cal A}$ and ${\cal B}$ 
satisfy the ortho-normalization condition and then form an $SO(2N+2)$ 
matrix ${\cal G}$ represented as  
\cite{Fuk.77}
\beq
{\cal G}
= 
\left[ 
\BA{cc}
{\cal A} & \bar{\cal B} \\
{\cal B} & \bar{\cal A}
\EA 
\right],               
~~~~
{\cal G}^{\dag } {\cal G}
=
{\cal G}{\cal G}^{\dag }
= 1_{2N+2} ,
\label{calG}
\eeq
which means
the ortho-normalization conditions of the $N+1$-dimensional HB amplitudes
\beqa
\left.
\BA{ll}
&
{\cal A}^\dag {\cal A} + {\cal B}^\dag {\cal B} 
= 
1_{N+1} ,~~
{\cal A}^{\mbox{\scriptsize T}}{\cal B} 
+ 
{\cal B}^{\mbox{\scriptsize T}}{\cal A} 
= 
0 ,\\
\\
&
{\cal A} {\cal A}^\dag 
+ 
\bar{\cal B} {\cal B}^{\mbox{\scriptsize T}} 
= 
1_{N+1} ,~~
\bar{\cal A} {\cal B}^{\mbox{\scriptsize T}} 
+ 
{\cal B}{\cal A}^\dag 
= 
0 .
\EA
\right\}
\label{HBorthonormalization}
\eeqa
The matrix ${\cal G}$ satisfies $\det {\cal G} = 1$ 
as is proved easily below
\beq
\det {\cal G}
=
\det
\left(
{\cal A} - \bar{\cal B} \bar{\cal A}^{-1} {\cal B}
\right)
\det \bar{\cal A}
=
\det
\left(
{\cal A}{\cal A}^\dag 
- 
\bar{\cal B} \bar{\cal A}^{-1} {\cal B}{\cal A}^\dag 
\right)
= 1 .
\label{detcalG}
\eeq
By using
(\ref{relAtoaXY}) and (\ref{defG}),
the matrices ${\cal A}$ and ${\cal B}$
can be decomposed as  
\beq
{\cal A}
\!=\! 
\left[ 
\BA{cc}
1_N - {\displaystyle \frac{\bar{x} r^{\mbox{\scriptsize T}}}{2}} & 
{\displaystyle -\frac{\bar{x}}{2}} \\
\\[-6pt]
{\displaystyle \frac{(1+z)r^{\mbox{\scriptsize T}}}{2}} & 
{\displaystyle \frac{1+z}{2}}
\EA 
\right] \!\!
\left[ 
\BA{cc}
a & 0 \\
\\ \\[-6pt]
0 & 1
\EA 
\right],
~
{\cal B}
\!=\! 
\left[ 
\BA{cc}
1_N + {\displaystyle \frac{x r^{\mbox{\scriptsize T}}q^{-1}}{2}} & 
{\displaystyle \frac{x}{2}} \\
\\[-6pt]
- {\displaystyle \frac{(1+z)r^{\mbox{\scriptsize T}}q^{-1}}{2}} & 
{\displaystyle \frac{1-z}{2}}
\EA 
\right] \!\!
\left[ 
\BA{cc}
b & 0 \\
\\ \\[-6pt]
0 & 1
\EA 
\right] ,
\label{calApcalBp}
\eeq
from which we get
the inverse of ${\cal A},~{\cal A}^{-1}$, as 
\beq
{\cal A}^{-1}
=
\left[ 
\BA{cc}
a^{-1} & 0 \\
\\
0 & 1
\EA 
\right]
\left[ 
\BA{cc}
1_N & {\displaystyle \frac{\bar{x}}{1+z}} \\
- r^{\mbox{\scriptsize T}} & 1
\EA 
\right] .
\label{calAinverse}
\eeq
From
(\ref{calApcalBp}) and (\ref{calAinverse}),
we obtain a $\frac{SO(2N+2)}{U(N+1)}$ coset variable 
with the $N$+1-th component 
as
\beq
{\cal Q}
=
{\cal B}{\cal A}^{-1}
= 
\left[ 
\BA{cc}
q & r \\
-r^{\mbox{\scriptsize T}} & 0
\EA 
\right]
=
-{\cal Q}^{\mbox{\scriptsize T}},
\label{cosetvariable2} 
\eeq
from which the $SO(2N+1)$ variables 
$q_{\alpha \beta }$ and $r_\alpha$
are shown to be just the independent variables of the
$\frac{SO(2N+2)}{U(N+1)}$ coset space. 
The paired mode $q_{\alpha \beta }$ and 
unpaired mode $r_\alpha$ variables in the $SO(2N+1)$ algebra 
are unified as the paired variables in the $SO(2N+2)$ algebra
\cite{Fuk.77}. 
We denote the $(N+1)$-dimension of the matrix $Q$ by
the index 0 and use the indices $p,~q,~\cdots$ 
running over 0 and $\alpha,~\beta,~\cdots$.


\newpage

\setcounter{equation}{0}
\renewcommand{\theequation}{\arabic{section}.\arabic{equation}}

\section{$\sigma$-model on the SO(2N+2)/U(N+1) coset manifold}
 
~~~
Let us introduce a $(2N+2) \times (N+1)$ isometric matrix ${\cal U}$ by
\beq
{\cal U}^{\mbox{\scriptsize T}}
=
\left[ 
\BA{cc} 
{\cal B}^{\mbox{\scriptsize T}}, ~ {\cal A}^{\mbox{\scriptsize T}}
\EA 
\right] .
\label{isomat}
\eeq
If one uses the matrix elements of 
${\cal U}$ and ${\cal U}^\dag $
as the co-ordinates on the manifold $SO(2N+2)$,
a real line element can be defined by a hermitian metric tensor 
on the manifold.
Under the transformation
${\cal U} \!\rightarrow\! {\cal VU}$
the metric is invariant.
Then the metric tensor defined on the manifold may become singular,
due to the fact that one uses too many co-ordinates.
 
According to Zumino
\cite{Zumino.79},
if ${\cal A}$ is non-singular,
we have relations governing ${\cal U}^\dag {\cal U}$ as
\beqa
\left.
\BA{ll}
&
{\cal U}^\dag {\cal U}
=
{\cal A}^\dag {\cal A}
+
{\cal B}^\dag {\cal B}
=
{\cal A}^\dag
\left\{
1_{N+1} 
+ 
\left(
{\cal B}{\cal A}^{-1} \right)^\dag \left({\cal B}{\cal A}^{-1} 
\right)
\right\}
{\cal A}
=
{\cal A}^\dag
\left(
1_{N+1} 
+ 
{\cal Q}^\dag {\cal Q}
\right)
{\cal A} ,\\
\\
&
\ln \det {\cal U}^\dag {\cal U}
=
\ln \det 
\left(
1_{N+1} 
+ 
{\cal Q}^\dag {\cal Q}
\right)
+
\ln \det {\cal A}
+
\ln \det {\cal A}^\dag ,
\EA
\right\}
\label{lndetUUdagger}
\eeqa
where we have used 
the $\frac{SO(2N+2)}{U(N+1)}$ coset variable ${\cal Q}$
(\ref{cosetvariable2}).
If we take the matrix elements of ${\cal Q}$ and $\bar{{\cal Q}}$ 
as the co-ordinates 
on the $\frac{SO(2N+2)}{U(N+1)}$ coset manifold,
the real line element can be well defined by a hermitian metric tensor 
on the coset manifold as
\beq
ds^2
=
{\cal G}_{pq}{~}_{\underline{r}\underline{s}}
d{\cal Q}^{pq}d\bar{{\cal Q}}^{\underline{r}\underline{s}}~
({\cal Q}^{pq} = {\cal Q}_{pq}~\mbox{and}~
\bar{{\cal Q}}^{\underline{r}\underline{s}} 
=
\bar{{\cal Q}}_{\underline{r}\underline{s}};~
{\cal G}_{pq}{~}_{\underline{r}\underline{s}}
=
{\cal G}_{\underline{r}\underline{s}}{~}_{pq}) .
\label{metric}
\eeq
We also use the indices
$\underline{r},~\underline{s},~\cdots$ 
running over 0 and $\alpha,~\beta,~\cdots$.
The condition that the manifold under consideration is 
a K\"{a}hler manifold 
is that its complex structure is 
covariantly constant relative to the Riemann connection:
\beq
{\cal G}_{pq~\underline{r}\underline{s}~,tu}
\stackrel{\mathrm{def}}{=}
\frac{\partial {\cal G}_{pq~\underline{r}\underline{s}}}
{\partial {\cal Q}^{tu}}
=
{\cal G}_{tu~\underline{r}\underline{s}~,pq} ,~~
{\cal G}_{pq~\underline{r}\underline{s}~,\underline{t}\underline{u}}
\stackrel{\mathrm{def}}{=}
\frac{\partial {\cal G}_{pq~\underline{r}\underline{s}}}
{\partial \bar{\cal Q}^{\underline{t}\underline{u}}}
=
{\cal G}_{pq~\underline{t}\underline{u}~,\underline{r}\underline{s}} ,
\label{Kcondition}
\eeq
and that it has vanishing torsions.
Then, the hermitian metric tensor 
${\cal G}_{pq}{~}_{\underline{r}\underline{s}}$
can be locally given through a real scalar function,
the K\"{a}hler potential, 
which takes the well-known form 
\beq
{\cal K}({\cal Q}^\dag,~{\cal Q}) 
=
\ln \det 
\left(
1_{N+1} 
+ 
{\cal Q}^\dag {\cal Q}
\right) ,
\label{Kaehlerpotential}
\eeq
and the explicit expression for the components of the metric tensor
is given as 
\beqa
\BA{ll}
{\cal G}_{pq~\underline{r}\underline{s}}
=
{\displaystyle 
\frac{\partial ^2 {\cal K}({\cal Q}^\dag,~{\cal Q})}
{\partial {\cal Q}^{pq} 
 \partial \bar{{\cal Q}}^{\underline{r}\underline{s}}}
}
&
=
\left\{
\left(
1_{N+1} 
+ 
{\cal Q}{\cal Q}^\dag 
\right)^{-1}
\right\}_{sp}
\left\{
\left(
1_{N+1} 
+ 
{\cal Q}^\dag {\cal Q}
\right)^{-1}
\right\}_{qr} \\
\\
&
~~~
-
(r \leftrightarrow s) - (p \leftrightarrow q) 
+
(p \leftrightarrow q,~r \leftrightarrow s) .
\EA
\label{metricfromKpot}
\eeqa
Notice that the above function does not determine
the K\"{a}hler potential 
${\cal K}({\cal Q}^\dag,~{\cal Q})$ 
uniquely
since the metric tensor 
${\cal G}_{pq}{}_{\underline{r}\underline{s}}$
is invariant under transformations of the K\"{a}hler potential,
\beq
{\cal K}({\cal Q}^\dag,~{\cal Q})
\rightarrow
{\cal K}^\prime ({\cal Q}^\dag,~{\cal Q})
=
{\cal K}({\cal Q}^\dag,~{\cal Q})
+ {\cal F}({\cal Q})
+ \bar{\cal F}(\bar{\cal Q}) .
\label{transKpot}
\eeq
${\cal F}({\cal Q})$ and $\bar{\cal F}(\bar{\cal Q})$
are analytic functions of ${\cal Q}$ and $\bar{\cal Q}$, respectively.
In the case of the K\"{a}hler metric tensor,
we have only the components of the metric connections 
with unmixed indices
\beq
\Gamma^{~tu}_{pq~rs}
=
{\cal G}^{\underline{v}\underline{w}~tu}
{\cal G}_{pq~\underline{v}\underline{w}~,rs} ,~~
\bar{\Gamma }^{~\underline{t}\underline{u}}_{\underline{p}\underline{q}~
\underline{r}\underline{s}}
=
{\cal G}^{\underline{t}\underline{u}~vw}
{\cal G}_{vw~\underline{r}\underline{s}~,\underline{p}\underline{q}} ,~~
{\cal G}^{\underline{v}\underline{w}~tu}
\stackrel{\mathrm{def}}{=}
({\cal G}^{-1})_{\underline{v}\underline{w}~tu} ,
\label{metricconect}
\eeq
and only the components of the curvatures
\beqa
\left.
\BA{ll}
&
\mbox{\boldmath $R$}_{pq~\underline{r}\underline{s}}
{~}_{tu~\underline{v}\underline{w}}
=
{\cal G}_{vw~\underline{v}\underline{w}}
\Gamma^{~vw}_{pq~tu}{}_{~, \underline{r}\underline{s}}
=
{\cal G}_{pq~\underline{v}\underline{w}}
{}_{~,tu~\underline{r}\underline{s}}
-
{\cal G}_{t^\prime u^\prime~\underline{v^\prime } \underline{w^\prime } }
\Gamma^{~t^\prime u^\prime~}_{pq~tu}
\bar{\Gamma }^{~\underline{v^\prime } \underline{w^\prime }}
_{\underline{r}\underline{s}~\underline{v}\underline{w}} ,\\
\\
&
\mbox{\boldmath $R$}_{\underline{r}\underline{s}~pq}
{~}_{\underline{v}\underline{w}~tu}
=
{\cal G}_{tu~\underline{t}\underline{u}}
\bar{\Gamma }^{~\underline{t}\underline{u}}
_{\underline{r}\underline{s}~\underline{v}\underline{w}}
{}_{~, pq} 
=
\mbox{\boldmath $R$}_{pq~\underline{r}\underline{s}} 
{~}_{tu~\underline{v}\underline{w}} .
\EA
\right\}
\label{curvature}
\eeqa

In two- or four-dimensional space-time,
the simplest representation of ${\cal N} \!\!=\!\! 1$ supersymmetry
is a scalar multiplet 
$\phi \!=\! \{{\cal Q}, \psi_L, H\}$
where ${\cal Q}$ and $H$ are complex scalars and
$
\psi_L
\!\equiv\!
\frac{1}{2}(1 + \gamma_5)\psi
$ 
is a left-handed chiral spinor
defined through a Majorana spinor.
In superspace language 
the multiplet is written as a chiral superfield:
\beq
\phi 
= 
{\cal Q} + \bar{\theta }_R \psi_L + \bar{\theta }_R \theta_L H .
\label{chiralsuperfield}
\eeq
The most general theory of the supersymmetric $\sigma$-model
can be constructed from the $[N]$ scalar multiplets
$\phi^{[\alpha]} 
\!=\! 
\{{\cal Q}^{[\alpha]},\psi^{[\alpha]}_L,H^{[\alpha]}\}
({[\alpha]} \!=\! 1,~\cdots,~[N])$.
The supersymmetry transformations are given by
\beq
\delta {\cal Q}^{[\alpha]}
=
\bar{\varepsilon }_R \psi^{[\alpha]}_L,~~
\delta \psi^{[\alpha]}_L
=
\frac{1}{2}
(
\slashed{\delta }{\cal Q}^{[\alpha]}\varepsilon_R 
+ 
H^{[\alpha]}\varepsilon_L
),~~
\delta H^{[\alpha]}
=
\bar{\varepsilon }_L \slashed{\delta }\psi^{[\alpha]}_L ,
\label{susytrans}
\eeq
where $\varepsilon$ is the Majorana spinor parameter.

Let the K\"{a}hler manifold be the $\frac{SO(2N+2)}{U(N+1)}$ coset manifold
and
redenote the complex scalar fields ${\cal Q}_{pq}$ 
as 
${\cal Q}^{[\alpha]}([\alpha] \!=\! 1,\cdots,~
\frac{N(N+1)}{2}~(\!=\! [N]))$
and
suppose that
the spinors $\psi_L ^{[\alpha]}$ and $\bar{\psi }_L ^{[\alpha]}$
span the fibres.
Following Zumino 
\cite{Zumino.79}
and van Holten et al.
\cite{vanHolten.85}, 
the Lagrangian of a supersymmetric $\sigma$-model 
can be written completely in terms of co-ordinates
on the fibre bundle in the following form:
\beqa
\BA{ll}
{\cal L}_{{\mbox{\scriptsize chiral}}} 
=
&\!\!\!\!
-
{\cal G}_{[\alpha][\underline{\beta }]}
\left(
\partial_\mu \bar{{\cal Q}}^{[\underline{\beta }]}
\partial_\mu {\cal Q}^{[\alpha]}
+
\bar{\psi }^{[\underline{\beta }]}_L
\overleftrightarrow{ \mbox{\boldmath $\slashed{D}$} }
\psi^{[\alpha]}_L
\right)
+
W_{;[\alpha][\beta]}
\bar{\psi }^{[\beta]}_R \psi^{[\alpha]}_L
+
\bar{W}_{;[\underline{\alpha }][\underline{\beta }]}
\bar{\psi }^{[\underline{\beta }]}_L \psi^{[\underline{\alpha }]}_R \\
\\
&\!\!\!\!
-
{\cal G}_{[\alpha][\underline{\alpha }]}
\bar{W}_{;[\underline{\alpha }]}
W_{;[\alpha]}
+
{\displaystyle \frac{1}{2}}
\mbox{\boldmath $R$}
_{[\alpha][\underline{\beta }][\gamma][\underline{\delta }]}
\bar{\psi }^{[\underline{\beta }]}_L \gamma_\mu \psi^{[\alpha ]}_L
\bar{\psi }^{[\underline{\delta }]}_L \gamma_\mu \psi^{[\gamma ]}_L ,
\EA
\label{susylaglangian}
\eeqa
and the K\"{a}hler covariant derivative is defined as
$
\mbox{\boldmath $D$}_\mu
\psi^{[\alpha ]}_L
\!\stackrel{\mathrm{def}}{=}\!
\partial_\mu \psi^{[\alpha ]}_L
+
\Gamma^{~[\alpha]}_{[\beta][\gamma]}
\psi^{[\beta]}_L
\partial_\mu {\cal Q}^{[\gamma]} 
$.
The curvature tensors 
$\mbox{\boldmath $R$}$
are given by
(\ref{curvature}).
This form of the Lagrangian has also been derived by Higashijima and Nitta
with the use of the K\"{a}hler normal co-ordinate expansion of
the Lagrangian 
$
{\cal L}
\!=\!
\int
d^4 \theta
{\cal K}(\phi^\dag,~\phi)
$
\cite{HN.01}.
The Lagrangian
${\cal L}_{{\mbox{\scriptsize chiral}}}$
(\ref{susylaglangian})
is manifestly a scalar under the general co-ordinate transformations
${\cal Q}^{[\alpha]} \!\rightarrow\! {\cal Q}^{\prime[\alpha]}$
on the manifold,
provided that
the fermion transforms as a vector and
the superpotential does as a scalar:
\beq
\psi^\prime{}^{[\alpha ]}_L
=
\frac{\partial {\cal Q}^{\prime[\alpha]}}
{\partial {\cal Q}^{[\beta]}}
\psi^{[\beta ]}_L,~~
W^\prime ({\cal Q}^\prime)
=
W({\cal Q}) .
\label{scalar}
\eeq
The auxiliary fields $H^{[\alpha]}$
are eliminated through their field equations
\beq
H^{[\alpha]}
=
\Gamma^{~[\alpha]}_{[\beta][\gamma]}
\bar{\psi }^{[\beta]}_R \psi^{[\gamma]}_L
+
{\cal G}^{[\alpha][\underline{\beta }]}
\bar{W}_{,[\underline{\beta }]} .
\label{auxilifield}
\eeq
The right-handed chiral spinor
$\psi_R$
is defined as 
$\psi_R
=
C\bar{\psi }^{\mbox{\scriptsize T}}_L$
and $C$ is the charge conjugation.
The comma
$,[\alpha]$ ($,[\underline{\alpha }]$) 
denotes a derivative with respect to
${\cal Q}^{[\alpha]}$ ($\bar{{\cal Q}}^{[\underline{\alpha }]}$),
while the semicolon denotes a covariant derivative
using the affine connection defined by 
the $\Gamma^{~[\alpha]}_{[\beta][\gamma]}$: 
\beq
W_{;[\alpha]}
=
W_{,[\alpha]}
=
\frac{\partial W}
{\partial {\cal Q}^{[\alpha]}},~~
W_{;[\alpha][\beta]}
=
\frac{\partial W_{;[\alpha]}}
{\partial {\cal Q}^{[\beta]}}
-
\Gamma^{~[\gamma]}_{[\beta][\alpha]}
W_{;[\gamma]} .
\label{covarideri}
\eeq



\newpage

\setcounter{equation}{0}
\renewcommand{\theequation}{\arabic{section}.\arabic{equation}}

\section{Expression for SO(2N+2)/U(N+1) Killing potential}
~~~Let us consider an $SO(2N+2)$ infinitesimal left transformation
of an $SO(2N+2)$ matrix ${\cal G}$ to ${\cal G}^\prime$,
$
{\cal G}^\prime
=
(1_{2N+2} + \delta {\cal G}) {\cal G}
$,
by using the first equation of
(\ref{infinitesimalop}):
\beqa
{\cal G}^\prime
= 
\left[ 
\BA{cc}
1_{N+1} + \delta {\cal A} & \delta \bar{\cal B} \\
\delta {\cal B} & 1_{N+1} + \delta \bar{\cal A}
\EA 
\right]
{\cal G}
=
\left[ 
\BA{cc}
{\cal A} + \delta {\cal A}{\cal A} + \delta \bar{\cal B} {\cal B}  & 
\bar{\cal B} + \delta {\cal A}\bar{\cal B} + 
\delta \bar{\cal B}\bar{\cal A} \\
{\cal B} + \delta \bar{\cal A} {\cal B} + \delta {\cal B}{\cal A}  & 
\bar{\cal A} + \delta \bar{\cal A}\bar{\cal A} + 
\delta {\cal B}\bar{\cal B}
\EA 
\right] .
\label{calGprime}
\eeqa
Let us define a $\frac{SO(2N+2)}{U(N+1)}$ coset variable
$
{\cal Q}^\prime
(\!=\! {\cal B}^\prime {\cal A}^{\prime -1})
$
in the ${\cal G}^\prime$ frame.
With the aid of 
(\ref{calGprime}),
the ${\cal Q}^\prime$ is calculated infinitesimally as
\beqa
\BA{ll}
{\cal Q}^\prime
=
{\cal B}^\prime {\cal A}^{\prime -1}
&=
\left(
{\cal B} + \delta \bar{\cal A} {\cal B} + \delta {\cal B}{\cal A}
\right)
\left(
{\cal A} + \delta {\cal A}{\cal A} + \delta \bar{\cal B} {\cal B}
\right)^{-1} \\
\\
&=
{\cal Q} + \delta {\cal B} 
- {\cal Q}\delta {\cal A} + \delta \bar{\cal A}{\cal Q}
- {\cal Q}\delta \bar{\cal B}{\cal Q} .
\EA
\label{calQprime}
\eeqa

The K\"{a}hler metrics admit a set of holomorphic isometries,
the Killing vectors,
${\cal R}^{i[\alpha]}({\cal Q})$
and
$\bar{\cal R}^{i[\underline{\alpha }]}(\bar{\cal Q})~
(i \!=\! 1, \cdots, \dim {\cal G})$,
which are the solution of the Killing equation
\beq
{\cal R}^i _{~[\underline{\beta }]}({\cal Q})_{,~[\alpha]}
+
\bar{\cal R}^i _{~[\alpha]}({\cal Q})_{,~[\underline{\beta }]}
=
0 ,~~
{\cal R}^i _{~[\underline{\beta }]}({\cal Q})
=
{\cal G}_{[\alpha][\underline{[\beta }]}{\cal R}^{i[\alpha]}({\cal Q}) .
\label{Killingeq}
\eeq
These isometries define infinitesimal symmetry transformations and 
are described geometrically by the above Killing vectors which
are the generators of infinitesimal co-ordinate transformations
keeping the metric invariant: 
$
\delta {\cal Q}
\!=\!
{\cal Q}^\prime \!-\! {\cal Q}
\!=\!
{\cal R}({\cal Q})
$
and
$
\delta \bar{\cal Q}
\!=\!
\bar{\cal R}(\bar{\cal Q})
$
such that
$
{\cal G}^\prime ({\cal Q}, \bar{\cal Q})
\!=\!
{\cal G} ({\cal Q}, \bar{\cal Q})
$.
The Killing equation
(\ref{Killingeq})
is the necessary and sufficient condition for 
an infinitesimal co-ordinate transformation
\beq
\delta{\cal Q}^{[\alpha]}
\!=\!
\left(
\delta {\cal B} 
- \delta {\cal A}^{\mbox{\scriptsize T}}{\cal Q} - {\cal Q}\delta {\cal A}
+ {\cal Q}\delta {\cal B}^\dag {\cal Q}
\right)^{[\alpha]}
\!=\!
\xi_i{\cal R}^{i[\alpha]}({\cal Q}) ,~~
\delta \bar{\cal Q}^{[\underline{\alpha }]}
\!=\!
\xi_i \bar{\cal R}^{i[\underline{\alpha }]}(\bar{\cal Q}) ,
\label{infinitesimaltrans}
\eeq
where 
$\xi _i$
are the infinitesimal and global group parameters.
Due to the Killing equation,
the Killing vectors
${\cal R}^{i[\alpha]}({\cal Q})$
and
$\bar{\cal R}^{i[\underline{\alpha }]}(\bar{\cal Q})$
can be written locally as the gradient of some real scalar function,
the Killing potentials
${\cal M}^i ({\cal Q}, \bar{\cal Q})$
such that
\beq
{\cal R}^i _{~[\underline{\alpha }]}({\cal Q})
=
-i{\cal M}^i _{~,[\underline{\alpha }]} ,~~
\bar{\cal R}^i _{~[\alpha]}(\bar{{\cal Q}})
=
i{\cal M}^i _{~,[\alpha]} .
\label{gradKillingpot}
\eeq

According to van Holten et al.
\cite{NNH.01}
and using the infinitesimal $SO(2N+2)$ matrix $\delta {\cal G}$
given by the first equation of
(\ref{infinitesimalop}),
the Killing potential ${\cal M}_\sigma$ 
can be written for the coset
$\frac{SO(2N+2)}{U(N+1)}$
as
\beqa
\left.
\BA{ll}
&
{\cal M}_\sigma
\left(
\delta {\cal A}, \delta {\cal B},\delta {\cal B}^\dag
\right)
=
\mbox{Tr}\left(\delta {\cal G} \widetilde{{\cal M}}_\sigma\right)
=
\mbox{tr}
\left(
\delta {\cal A} {\cal M}_{\sigma \delta {\cal A}}
+
\delta {\cal B} {\cal M}_{\sigma \delta {\cal B}^\dag }
+
\delta {\cal B}^\dag {\cal M}_{\sigma \delta {\cal B}}
\right) ,\\
\\
&
\widetilde{{\cal M}}_\sigma
\equiv
\left[ 
\BA{cc} 
 \widetilde{{\cal M}}_{\sigma \delta {\cal A}} & 
 \widetilde{{\cal M}}_{\sigma \delta {\cal B}^\dag }\\
 \\
-\widetilde{{\cal M}}_{\sigma \delta {\cal B}} & 
-\widetilde{{\cal M}}_{\sigma \delta {\cal A}^{\mbox{\scriptsize T}}} 
\EA 
\right] ,~~
\BA{c}
{\cal M}_{\sigma \delta {\cal A}}
=
\widetilde{{\cal M}}_{\sigma \delta {\cal A}}
+
\left(
\widetilde{{\cal M}}_{\sigma \delta {\cal A}^{\mbox{\scriptsize T}}}
\right)^{\mbox{\scriptsize T}} ,\\
\\
{\cal M}_{\sigma \delta {\cal B}}
=
\widetilde{{\cal M}}_{\sigma \delta {\cal B}} ,~~
{\cal M}_{\sigma \delta {\cal B}^\dag }
=
\widetilde{{\cal M}}_{\sigma \delta {\cal B}^\dag } ,
\EA
\EA
\right\}
\label{KillingpotM}
\eeqa
where the trace Tr is taken over the $(2N+2) \!\times\! (2N+2)$ matrices,
while the trace tr is taken over the $(N+1) \!\times\! (N+1)$ matrices.
Let us introduce the $(N+1)$-dimensional matrices 
${\cal R}({\cal Q}; \delta {\cal G})$, 
${\cal R}_T({\cal Q}; \delta {\cal G})$ and ${\cal X}$ by
\beqa
\left.
\BA{ll}
&
{\cal R}({\cal Q}; \delta {\cal G})
\!=\!
\delta {\cal B} 
- \delta {\cal A}^{\mbox{\scriptsize T}}{\cal Q} - {\cal Q}\delta {\cal A}
+ {\cal Q}\delta {\cal B}^\dag {\cal Q} ,~~
{\cal R}_T ({\cal Q}; \delta {\cal G})
\!=\!
-\delta {\cal A}^{\mbox{\scriptsize T}}
+ 
{\cal Q}\delta {\cal B}^\dag ,\\
\\
&
{\cal X}
=
(1_{N+1} + {\cal Q}{\cal Q}^\dag)^{-1}
= 
{\mathcal X}^\dag .
\EA
\right\}
\label{RRTChi}
\eeqa
In 
(\ref{infinitesimaltrans}),
putting $\xi_i \!=\! 1$,
we have
$\delta {\cal Q} 
\!=\! 
{\cal R}({\cal Q}; \delta {\cal G})
$ 
which is just the Killing vector
in the coset space
$\frac{SO(2N+2)}{U(N+1)}$,
and tr of the holomorphic matrix-valued function 
${\cal R}_T ({\cal Q}; \delta {\cal G})$,
$\mbox{tr}{\cal R}_T ({\cal Q}; \delta {\cal G}) \!=\! {\cal F}({\cal Q})$
is a holomorphic K\"{a}hler transformation. 
Then the Killing potential ${\cal M}_\sigma$ is given as
\beqa
\!\!\!\!\!\!\!\!
\left.
\BA{rl}
&-i{\cal M}_\sigma
\left(
{\cal Q}, \bar{\cal Q};\delta {\cal G}
\right)
=
-\mbox{tr}
\Delta
\left(
{\cal Q}, \bar{\cal Q};\delta {\cal G}
\right) ,\\
\\
&\Delta
\left(
{\cal Q}, \bar{\cal Q};\delta {\cal G}
\right)
\stackrel{\mathrm{def}}{=}
{\cal R}_T ({\cal Q}; \delta {\cal G})
-
{\cal R}({\cal Q}; \delta {\cal G}) {\cal Q}^\dag {\cal X} 
=
\left(
{\cal Q} \delta {\cal A} {\cal Q}^\dag 
- 
\delta {\cal A}^{\mbox{\scriptsize T}}
-
\delta {\cal B} {\cal Q}^\dag 
+
{\cal Q} \delta {\cal B}^\dag
\right)
{\cal X} .
\EA \!\!
\right\}
\label{formKillingpotM} 
\eeqa
From
(\ref{KillingpotM}) and (\ref{formKillingpotM}),
we obtain
\beq
-i{\cal M}_{\sigma \delta {\cal B}}
=
-{\cal X} {\cal Q} ,~~
-i{\cal M}_{\sigma \delta {\cal B}^\dag }
=
 {\cal Q}^\dag {\cal X} ,~~
-i{\cal M}_{\sigma \delta {\cal A}}
=
1_{N+1} - 2 {\cal Q}^\dag {\cal X} {\cal Q} .
\label{componentKillingpotM} 
\eeq
Using the expression for $\widetilde{{\cal M}}_\sigma$,
equation
(\ref{componentKillingpotM}),
their components are written in the form
\beq
-i\widetilde{{\cal M}}_{\! \sigma \delta {\cal B}}
\!=\!
-{\cal X} \! {\cal Q} ,~
-i\widetilde{{\cal M}}_{\! \sigma \delta {\cal B}^\dag }
\!=\!
 {\cal Q}^\dag \! {\cal X} ,~
-i\widetilde{{\cal M}}_{\! \sigma \delta {\cal A}}
\!=\!
-{\cal Q}^\dag \! {\cal X} \! {\cal Q} ,~
-i\widetilde{{\cal M}}_{\! \sigma \delta {\cal A}^{\mbox{\scriptsize T}}}
\!=\!
1_{\! N \!+\! 1} - {\cal Q} \! \bar{\cal X} \! {\cal Q}^\dag
\!=\!
{\cal X} .
\label{tildecomponentKillingpotM} 
\eeq
It is easily checked that
the result
(\ref{componentKillingpotM})
satisfies the gradient of the real function ${\cal M}_{\sigma }$
(\ref{gradKillingpot}).
Of course, putting $r \!=\! 0$ in ${\cal Q}$
(\ref{cosetvariable2}),
the Killing potential ${\cal M}_\sigma$
in the $\frac{SO(2N+2)}{U(N+1)}$ coset space
leads to 
the Killing potential $M_\sigma$
in the $\frac{SO(2N)}{U(N)}$ coset space
obtained by van Holten et al.
\cite{NNH.01}.

To make clear the meaning of the Killing potential,
using the $(2N+2) \!\times\! (N+1)$ isometric matrix 
${\cal U}~({\cal U}^\dag{\cal U}\!=\!1_{N+1})$,
let us introduce the following $(2N+2) \!\times\! (2N+2)$ matrix:
\beq
{\cal W}
=
{\cal U}{\cal U}^\dag
=
\left[ 
\BA{cc} 
{\cal R} & {\cal K} \\
\\[-8pt]
-\bar{\cal K} & 1_{N+1} - \bar{\cal R}
\EA 
\right] ,~~
\BA{c}
{\cal R}
=
{\cal B}{\cal B}^\dag ,\\
\\[-8pt]
{\cal K}
=
{\cal B}{\cal A}^\dag  ,
\EA
\label{densitymat}
\eeq
which satisfies the idempotency relation 
${\cal W}^2 \!=\! {\cal W}$
and is hermitian
on the $SO(2N+2)$ group.
${\cal W}$ is a natural extension of 
the generalized density matrix in the $SO(2N)$ CS rep
to the $SO(2N+2)$ CS rep.
Since the matrices ${\cal A}$ and ${\cal B}$
are represented in terms of ${\cal Q}\!=\!({\cal Q}_{pq})$ as
\beq
{\cal A}
=
(1_{N+1} + {\cal Q}^\dag{\cal Q})^{-\frac{1}{2}}
\stackrel{\circ}{{\cal U}} ,~~
{\cal B}
=
{\cal Q}
(1_{N+1} + {\cal Q}^\dag{\cal Q})^{-\frac{1}{2}}
\stackrel{\circ}{{\cal U}} ,~~
\stackrel{\circ}{{\cal U}} \in U(N+1) ,
\label{matAandB}
\eeq
then, we have
\beq
{\cal R}
=
{\cal Q}(1_{N+1} + {\cal Q}^\dag{\cal Q})^{-1}{\cal Q}^\dag 
=
{\cal Q}\bar{\chi }{\cal Q}^\dag
=
1_{N+1} - \chi ,~~
{\cal K}
=
{\cal Q}
(1_{N+1} + {\cal Q}^\dag{\cal Q})^{-1}
=
\chi {\cal Q}
\label{matRandK}
\eeq
Substituting
(\ref{matRandK})
into
(\ref{tildecomponentKillingpotM}),
the Killing potential $-i\widetilde{{\cal M}}_\sigma$
is expressed in terms of the sub-matrices ${\cal R}$ and ${\cal K}$ of 
the generalized density matrix
(\ref{densitymat}) 
as
\beq
-i\widetilde{{\cal M}}_\sigma
=
\left[ 
\BA{cc} 
-\bar{\cal R} & -\bar{\cal K} \\
\\[-8pt]
{\cal K} & -(1_{N+1} - {\cal R})
\EA 
\right] ,
\label{tildecomponentKillingpotW}
\eeq
from which
we finally obtain
\beq
-i\overline{\widetilde{{\cal M}}}_\sigma
=
\left[ 
\BA{cc} 
{\cal R} & {\cal K} \\
\\[-8pt]
-\bar{\cal K} & 1_{N+1} - \bar{\cal R}
\EA 
\right] .
\label{complextildecomponentKillingpotW}
\eeq
To our great surprise,
the expression for the Killing potential
(\ref{complextildecomponentKillingpotW})
just becomes equivalent with
the generalized density matrix
(\ref{densitymat}).

The expression for the Killing potential ${\cal M}_\sigma$
is described in terms of
the $\frac{SO(2N+2)}{U(N+1)}$ coset variable 
${\cal Q}_{pq}$
but include an inverse matrix 
${\cal X}$
given by
(\ref{RRTChi}).
The variable ${\cal Q}_{pq}$ has already been expressed
in terms of the variables $q_{\alpha \beta }$ and $r_\alpha$
through
(\ref{cosetvariable2}).
To obtain the concrete expression for the Killing potential
in terms of the $SO(2N+1)$ variables $q_{\alpha \beta }$ and $r_\alpha$,
we must also represent the inverse matrix
in terms of the variables $q_{\alpha \beta }$ and $r_\alpha$.
Following Ref.
\cite{Ni.98}, 
after some algebraic manipulations, 
the inverse matrix 
${\cal X}$
in (\ref{RRTChi}) 
leads to the form
\beq
{\mathcal X}
=
\left[ 
\BA{cc}
{\cal Q}_{qq^\dag } & {\cal Q}_{q r} \\
{\cal Q}_{q r}^{\dag } & {\cal Q}_{r^\dag r}
\EA 
\right] ,~~
\chi
=
(1_N + q q^\dag )^{-1} 
=
\chi^\dag ,
\label{inverse1plusQQ}
\eeq
where each sub-matrix is expressed by the variables $q$ and $r$ as
\beq
{\cal Q}_{qq^\dag }
=
\chi -\frac{1+z}{2}\chi(r r^\dag  
- q\bar{r}r^{\mbox{\scriptsize T}}q^\dag )
\chi,
\label{inverse1plusQQ1}
\eeq
\beq
{\cal Q}_{q \bar{r}}
=
\frac{1+z}{2}\chi q \bar{r} ,~~~
{\cal Q}_{r^\dag r}
=
\frac{1+z}{2} .
\label{inverse1plusQQsub23}
\eeq

Then, substituting 
(\ref{cosetvariable2}) and (\ref{inverse1plusQQ})
into
(\ref{componentKillingpotM})
and introducing an auxiliary function
$
\lambda
\!=\!
r r^\dag  \!-\! q\bar{r}r^{\mbox{\scriptsize T}}q^\dag
\!=\!
\lambda^\dag
$,
we can get the Killing potential ${\cal M}_\sigma$
expressed in terms of only $q,~r$ and $1 + z \!=\! 2Z^2$
as,

\beq
-i{\cal M}_{\sigma \delta {\cal B}}
=
\left[ 
\BA{cc}
-
\chi q 
\!+\! 
Z^2
\left(
\chi \lambda \chi q
\!+\!
\chi q \bar{r} r^{\mbox{\scriptsize T}}
\right)  &
-
\chi r 
\!+\! 
Z^2
\chi \lambda \chi r  \\
\\
-
Z^2
\left(
r^{\mbox{\scriptsize T}} q^\dag \chi q
\!-\!
r^{\mbox{\scriptsize T}}
\right) &
- 
Z^2
r^{\mbox{\scriptsize T}} q^\dag \chi r 
\EA 
\right] ,
\label{KillingpotB} 
\eeq
\vspace{0.5cm}
\beq
-i{\cal M}_{\sigma \delta {\cal B}^\dag }
=
\left[ 
\BA{cc}
q^\dag \chi 
\!-\! 
Z^2
\left(
q^\dag \chi \lambda \chi 
\!\!+
\bar{r} r^{\mbox{\scriptsize T}} q^\dag \chi
\right)  &
Z^2
\left(
q^\dag \chi q \bar{r}
\!-\!
\bar{r}
\right) \\
\\
r ^\dag \chi 
\!-\! 
Z^2
r ^\dag \chi \lambda \chi & 
Z^2
r^\dag  \chi q \bar{r} 
\EA 
\right] ,
\label{KillingpotBdag}
\eeq

\beqa
\!\!\!\!\!\!\!\!\!\!\!\!\!
\BA{ll}
&~
-i{\cal M}_{\sigma \delta {\cal A}} = \\
\\
&
\left[ \!\!\!\!
\BA{cc}
1_N \!-\! 2q^\dag \chi q 
\!+\! 
2Z^2 \!
\left( \!
q^\dag \chi \lambda \chi q
\!+\!
q^\dag \chi q \bar{r} r^{\mbox{\scriptsize T}}
\!+\!
\bar{r} r^{\mbox{\scriptsize T}} q^\dag \chi q
\!-\!
\bar{r} r^{\mbox{\scriptsize T}} \!
\right) \!\! & 
\!-\! 
2 q^\dag \chi r 
\!+\! 
2Z^2 \!
\left(
q^\dag \chi \lambda \chi r
\!+\!
\bar{r} r^{\mbox{\scriptsize T}} \! q^\dag \chi r
\right) \!\!\!\! \\
\\
\!-\! 2r^\dag \chi q 
\!+\! 
2Z^2 \!
\left( \!
r^\dag \chi \lambda \chi q
\!+\!
r^\dag \chi q \bar{r} r^{\mbox{\scriptsize T}} \!
\right) &
1
\!-\!
2 r^\dag \chi r 
\!+\! 
2Z^2 \!
r^\dag \chi \lambda \chi r \!\!\!
\EA 
\right] \!\!.
\EA
\label{KillingpotA} 
\eeqa
In the above expressions for the Killing potential ${\cal M}_\sigma$,
each block-matrix is easily verified to satisfy 
the following identities and relations:
\beq
r^{\mbox{\scriptsize T}} q^\dag \chi r
=
0 ,~~
r ^\dag \chi q \bar{r}
=
0 ,~~
r^\dag \chi r
=
\frac{1- Z^2}{Z^2} ,~~
r^\dag \chi \lambda \chi r
=
\left(
\frac{1- Z^2}{Z^2} 
\right)^2 ,
\label{identities} 
\eeq
\vspace{0.1cm}
\beq
1
-
2 r^\dag \chi r 
+ 
2Z^2 
r^\dag \chi \lambda \chi r
=
2Z^2 -1 ,
\label{relation1}
\eeq
\vspace{0.1cm}
\beq
\chi \lambda \chi r
=
\frac{1- Z^2}{Z^2} \chi r ,~~
r^\dag \chi \lambda \chi 
=
\frac{1- Z^2}{Z^2} r^\dag \chi ,~~
q^\dag \chi q
=
1_N - \bar{\chi } .
\label{relation2}
\eeq
Using these identities and relations,
we get compact forms of the Killing potential ${\cal M}_\sigma$
as,
\beqa
-i{\cal M}_{ 
\BA{c}
{\scriptstyle \sigma \delta {\cal B}} \\
{\scriptstyle \left(\sigma \delta {\cal B}^\dag \right)}
\EA
}
=
\left[ 
\BA{cc}
\BA{c}
-
\chi q 
+ 
Z^2
\left(
\chi r r^\dag \chi q
+
\chi q \bar{r} r^{\mbox{\scriptsize T}} \bar{\chi }
\right) \\
\left(
q^\dag \chi 
- 
Z^2
\left(
q^\dag \chi r r^\dag \chi 
+
\bar{\chi } \bar{r} r^{\mbox{\scriptsize T}} q^\dag \chi
\right)
\right)
\EA &
\BA{c} 
- Z^2 \chi r \\
\left(
- Z^2 \bar{\chi } \bar{r}
\right)
\EA \\
\\
\BA{c}
Z^2
r^{\mbox{\scriptsize T}} \bar{\chi } \\
\left(
Z^2
r^\dag \chi
\right)
\EA 
&
\BA{c} 
0 \\ 
\left(
0
\right)
\EA
\EA 
\right] ,
\label{KillingpotB2Bdag2} 
\eeqa
\beq
-i{\cal M}_{\sigma \delta {\cal A}} 
= 
\left[ 
\BA{cc}
1_N - 2q^\dag \chi q 
+ 
2Z^2 
\left( 
q^\dag \chi r r^\dag \chi q
-
\bar{\chi } \bar{r} r^{\mbox{\scriptsize T}} \bar{\chi }
\right)  & - 2Z^2 q^\dag \chi r \\
\\
- 2Z^2 
r^\dag \chi q & 2Z^2 - 1
\EA 
\right] .
\label{KillingpotA2}
\eeq

Let us introduce the gauge covariant derivatives
\beqa
\left.
\BA{rl}
&
\mbox{\boldmath $D$}_\mu {\cal Q}^{[\alpha]}
=
\partial_\mu {\cal Q}^{[\alpha]}
-
g_i A^i _{~\mu } {\cal R}^{i[\alpha]}({\cal Q}) ,\\
\\
&
\mbox{\boldmath $D$}_\mu
\psi^{[\alpha ]}_L
=
\partial_\mu \psi^{[\alpha ]}_L
-
g_i A^i _{~\mu } {\cal R}^{i[\alpha]} _{~~~,[\beta]}({\cal Q})\psi^{[\beta]}_L
+
\Gamma^{~[\alpha]}_{[\beta][\gamma]}
\psi^{[\beta]}_L
\partial_\mu {\cal Q}^{[\gamma]} ,
\EA
\right\}
\label{covariantders}
\eeqa 
where $A^i _{~\mu }$
are gauge fields corresponding to local symmetries and
$g_i$ are coupling constants.
They are components of vector multiplets
$V^i \!=\! (A^i _{~\mu }, \lambda^i, D^i)$,
with $\lambda^i$ representing the gauginos and
$D^i$ the real auxiliary fields.
With the introduction of the gauge fields in Lagrangian
(\ref{susylaglangian}),
via the gauge covariant derivatives
(\ref{covariantders}),
the $\sigma$-model is no longer invariant
under the supersymmetry transformations.
To restore the supersymmetry,
it is necessary to add the terms
\beq
\Delta {\cal L}_{\mbox{{\scriptsize chiral}}}
=
2 {\cal G}_{[\alpha][\underline{\alpha }]} 
\left( 
{\cal R}^i _{~[\underline{\alpha }]}({\cal Q}) 
\bar{\psi }^{[\underline{\alpha }]}_L \lambda^i _R
+
\bar{\cal R}^i _{~[\alpha]}({\cal Q}) 
\bar{\lambda }^i _R \psi^{[\alpha]}_L 
\right)
-
g_i \mbox{tr}
\left\{D^i ({\cal M}^i + \xi^i)\right\} ,
\label{addingterm}
\eeq
where $\xi_i$ are Fayet-Ilipoulos parameters. 
Then the full Lagrangian for this model consists of 
the usual supersymmetry Yang-Mills part and the chiral part
\beq
{\cal L}
=
-
\mbox{tr}
\left\{
\frac{1}{4}{\cal F}^i _{\mu \nu }{\cal F}^i _{\mu \nu }
+
\frac{1}{2}\bar{\lambda }^i \slashed{D} \lambda^i
-
\frac{1}{2}D^{i}D^{i}
\right\}
+
{\cal L}_{{\mbox{\scriptsize chiral}}} 
(\partial_\mu \rightarrow \mbox{\boldmath $D$}_\mu)
+ 
\Delta {\cal L}_{\mbox{{\scriptsize chiral}}} .
\label{fullLaglangian}
\eeq
Eliminating the auxiliary field $D^i$ by 
$
D^i
\!=\!
-g_i ({\cal M}^i \!+\! \xi^i)
$
(not summed for $i$),
we can get a scalar potential
\beq
V_{{\mbox{\scriptsize SC}}}
=
-
\frac{1}{2} 
g^2 _i \mbox{tr}
\left\{({\cal M}^i + \xi^i)^2\right\} ,
\label{scalerpot}
\eeq
in which a reduced scalar potential arising from the gauging of 
$SU(N+1) \!\times\! U(1)$ including 
a Fayet-Ilipoulos term with parameter $\xi$ is of special interest:
\beq
V_{{\mbox{\scriptsize redSC}}}
=
\frac{g^2 _{U(1)}}{2(N+1)}
\left(
\xi - i {\cal M}_Y
\right)^2
+
\frac{g^2 _{SU(N+1)}}{2}
\mbox{tr}
\left(
- i {\cal M}_t
\right)^2 .
\label{specialscalerpot} 
\eeq
The new quantities
$
\mbox{tr}
\left(
- i {\cal M}_t
\right)^2
$ 
and 
$- i {\cal M}_Y$
in $V_{{\mbox{\scriptsize redSC}}}$
are defined below
and the trace {\bf tr}, taken over the $N \!\times\! N$ matrix,
is used.
Then we have
\beqa
\left.
\BA{rl}
\mbox{tr}
(- i {\cal M}_t )^2
=
&\!\!\!\!
\mbox{tr}
(- i {\cal M}_{\sigma \delta {\cal A}})^2
-
{\displaystyle \frac{1}{N+1}}
(- i {\cal M}_Y)^2 ,~~
- i {\cal M}_Y
=
\mbox{tr}
\left(
- i {\cal M}_{\sigma \delta {\cal A}} 
\right), \\
\\[-6pt]
\mbox{tr}
\left(
-i{\cal M}_{\sigma \delta {\cal A}}
\right)
=
&\!\!\!\!
- N 
+ 
2 \mbox{{\bf tr}}(\chi)
+
2 Z^2 \mbox{{\bf tr}}(\chi r r^\dag)
-
4 Z^2 \mbox{{\bf tr}}(\chi r r^\dag \chi)
+
2 Z^2 - 1 , \\
\\[-6pt]
\mbox{tr}
\left(
-i{\cal M}_{\sigma \delta {\cal A}}
\right)^2
=
&\!\!\!\!
N 
- 
4 \mbox{{\bf tr}}(\chi )
+
4 \mbox{{\bf tr}}(\chi \chi)
+
12 Z^2 \mbox{{\bf tr}}(\chi r r^\dag \chi)
-
16 Z^2 \mbox{{\bf tr}}(\chi \chi r r^\dag \chi) \\
\\[-6pt]
&\!\!\!\!
-
4 Z^4 r^\dag \chi \chi r \cdot \mbox{{\bf tr}}(\chi r r^\dag )
+
8 Z^4 r^\dag \chi \chi r \cdot \mbox{{\bf tr}}(\chi r r^\dag \chi)
+
1 - 4 Z^4 r^\dag \chi \chi r ,
\EA
\right\}
\label{trM2}
\eeqa

Here we give the proof of the third identity of 
(\ref{identities})
as
\beqa
\BA{rl}
r^\dag \chi r
=
&\!\!\!\!
{\displaystyle 
\frac{1}{4Z^4}
(x^\dag + x^{\mbox{\scriptsize T}} q^\dag ) \chi (x + q \bar{x})
=
\frac{1}{4Z^4}
(x^\dag \chi x + x^{\mbox{\scriptsize T}} q^\dag \chi q \bar{x})
} \\
\\
=
&\!\!\!\!
{\displaystyle 
\frac{1}{4Z^4}
x^{\mbox{\scriptsize T}} \bar{x}
=
\frac{1 - z^2}{4Z^4}
=
\frac{1 - Z^2}{Z^2}
} .
\EA
\label{rdagchir}
\eeqa
By using the same method as the above,
we can approximately calculate the quantities
$r^\dag  \chi \chi r$ and $\mbox{{\bf tr}}(rr^\dag )$  
as
\beqa
\left.
\BA{rl}
r^\dag  \chi \chi r
=
&\!\!\!\!
{\displaystyle
\frac{1}{4Z^4}
(x^\dag + x^{\mbox{\scriptsize T}} q^\dag) \chi \chi (x + q \bar{x})
=
\frac{1}{4Z^4}
x^\dag \chi x 
} \\
\\[-8pt]
\approx
&\!\!\!\!
{\displaystyle \frac{1}{4Z^4}}
\left\{
\frac{1}{N}
\left[
N + \mbox{{\bf tr}}(q^\dag q)
\right]
\right\}^{-1}
x^\dag x
=
{\displaystyle \frac{1 - Z^2}{Z^2}}
< \!\chi\! >, \\
\\[-8pt]
< \!\chi\! >
\stackrel{\mathrm{def}}{=}
&\!\!\!\!
\left\{
{\displaystyle \frac{1}{N}}
\left[
N + \mbox{{\bf tr}}(q^\dag q)
\right]
\right\}^{-1} ,
\EA
\right\}
\label{rdagchichir}
\eeqa
\vspace{-0.5cm}
\beqa
\BA{rl}
\mbox{{\bf tr}}(rr^\dag )
=
&\!\!\!\!
r^\dag r
=
{\displaystyle
\frac{1}{4Z^4}
(x^\dag + x^{\mbox{\scriptsize T}} q^\dag) (x + q \bar{x})
=
\frac{1}{4Z^4}
x^\dag \chi^{-1} x 
} \\
\\[-8pt]
\approx
&\!\!\!\!
{\displaystyle
\frac{1 - Z^2}{Z^2}
\frac{1}{< \!\chi\! >}
} 
\stackrel{\mathrm{def}}{=}
< \!r r^\dag\! > .
\EA
\label{trrrdag}
\eeqa
In
(\ref{trM2}),
approximating
$\mbox{{\bf tr}}(\chi )$,
$\mbox{{\bf tr}}(\chi rr^\dag )$, etc.
by
$< \!\!\chi\!\! >$,
$< \!\!\chi\!\! > \! \mbox{{\bf tr}}(rr^\dag )$, etc.,
respectively,
and using
(\ref{rdagchichir}) and (\ref{trrrdag}),
$
\mbox{tr}
\left(
-i{\cal M}_{\sigma \delta {\cal A}}
\right)
$ 
and 
$
\mbox{tr}
\left(
-i{\cal M}_{\sigma \delta {\cal A}}
\right)^2
$
are computed as
\beqa
\left.
\BA{rl}
\mbox{tr}
\left(
-i{\cal M}_{\sigma \delta {\cal A}}
\right)
=
&\!\!\!\!
1 - N + 2(2Z^2 \!-\! 1) \! < \!\chi\! > , \\
\\
\mbox{tr}
\left(
-i{\cal M}_{\sigma \delta {\cal A}}
\right)^2
=
&\!\!\!\!
1 + N - 4(2Z^2 \!-\! 1) \! < \!\chi\! > \! 
+ 
4(2Z^4 \!-\! 1) \!  < \!\chi\! >^2 .
\EA
\right\}
\label{approxKillingpot}
\eeqa
Substituting
(\ref{approxKillingpot})
into
(\ref{specialscalerpot}),
we obtain the final form of the reduced scalar potential
as
\beqa
\BA{rl}
V_{{\mbox{\scriptsize redSC}}}
=
&\!\!\!\!
{\displaystyle
\frac{g^2 _{U(1)}}{2(N \!+\! 1)}
}
\left\{
\xi \!+\! 1 \!-\! N + 2(2Z^2 \!-\! 1) \! < \!\chi\! >
\right\}^2 \\
\\[-12pt]
&\!\!\!\!
+
{\displaystyle
2 
\frac{g^2 _{SU(N \!+\! 1)}}{N \!+\! 1}
}
\left[
N \!-\! 2(2Z^2 \!-\! 1) \! < \!\chi\! > 
\!+\!
\left\{
2(N \!-\! 1) Z^4 \!+\! 4 Z^2 \!-\! (N \!+\! 2)
\right\} \!
< \!\chi\! >^2
\right] ,
\EA
\label{specialscalerpot2} 
\eeqa
which is written in terms of the $SO(2N \!+\! 1)$ parameter $Z$,
the mean value of $\chi$, i.e., $< \!\!\chi\!\! >$, and
the Fayet-Ilipoulos parameter $\xi$.

In order to see the behaviour of the vacuum expectation value of 
the $\sigma$-fields,
it is very important to analyze the form of the reduced scalar potential.
From the variation of the reduced scalar potential
with respect to $Z$ and $< \!\!\chi\!\! >$,
we can obtain the following relations:
\beq
g^2 _{U(1)}
\left\{
\xi \!+\! 1 \!-\! N + 2(2Z^2 \!-\! 1) \! < \!\chi\! >
\right\}
-
2 
g^2 _{SU(N \!+\! 1)}
\left\{
1 - ((N - 1) Z^2 + 1) < \!\chi\! >
\right\}
= 0 ,
\label{VvariZ} 
\eeq
\beqa
\BA{rl}
&
g^2 _{U(1)}
\left\{
\xi \!+\! 1 \!-\! N + 2(2Z^2 \!-\! 1) \! < \!\chi\! >
\right\}
(2Z^2 \!-\! 1) \\
\\
&
-
2 
g^2 _{SU(N \!+\! 1)}
\left[
2Z^2 \!-\! 1
-
\left\{
2(N - 1) Z^4 + 4Z^2 - (N + 2)
\right\}
< \!\chi\! >
\right]
= 0 . 
\EA
\label{Vvarichi}
\eeqa
Multiplying by $2 Z^2 \!-\! 1$ for
(\ref{VvariZ} ) 
and using 
(\ref{Vvarichi}),
we have a $g^2$-independent relation
\beqa
\BA{rl}
&
\left\{
1 \!-\! ((N \!-\! 1)Z^2 \!+\! 1) \! < \!\chi\! >
\right\}
(2Z^2 \!-\! 1) \\
\\
&
-
\left[
2Z^2 \!-\! 1
-
\left\{
2(N - 1) Z^4 + 4Z^2 - (N + 2)
\right\}
< \!\chi\! > 
\right]
= 0 .
\EA
\label{eqdetermineZ} 
\eeqa
This relation reads
\beq
(N + 1)(Z^2 - 1) < \!\chi\! > = 0 ,
\label{simplesolutionforz2}
\eeq
from which, since $< \!\!\chi\!\! > \neq 0$, 
we get a very simple solution 
\beq
Z^2 = 1,~~
< \!\chi\! > 
= 
\frac{1}{2}
\frac{1}{g^2 _{U(1)} + N g^2 _{SU(N \!+\! 1)}}
\left\{
g^2 _{U(1)} (N - 1) + 2 g^2 _{SU(N \!+\! 1)}
-
g^2 _{U(1)} \xi
\right\} .
\label{simplesolutionforchi}
\eeq
This solution 
just corresponds to the 
$\frac{SO(2N)}{U(N)}$ supersymmetric $\sigma$-model
since $Z^2 \!=\! 1$.
Putting this solution into
(\ref{specialscalerpot2}),
the minimization of 
the reduced scalar potential
with respect to 
the Fayet-Ilipoulos parameter $\xi$
is realized
as follows:
\beqa
\!\!\!\!
\left.
\BA{rl}
&\!\!\!\!
V_{{\mbox{\scriptsize redSC}}}
=
{\displaystyle \frac{1}{2}}
{\displaystyle \frac{N}{N \!+\! 1}}
{\displaystyle
\frac{g^2 _{U(1)}g^2 _{SU(N + 1)}}{g^2 _{U(1)} \!+\! N g^2 _{SU(N + 1)}}
}
\left[
\xi 
\!+\!
{\displaystyle \frac{1}{N}} 
\left\{
2
-
N
\left( N \!-\! 1 \right)
\right\}
\right]^2 
\!+\!
V_{{\mbox{\scriptsize redSC}}}^{\mbox{\scriptsize min}} ,
\\
\\[-10pt]
&\!\!\!\!
~~~
\xi_{\mbox{\scriptsize min}} 
=
N
\!-\!
1
\!-\! 
2 
{\displaystyle \frac{1}{N}} ,~~
V_{{\mbox{\scriptsize redSC}}}^{\mbox{\scriptsize min}}
\!=\!
{\displaystyle
2 
\frac{g^2 _{SU(N + 1)}}{N + 1}
}
\left(
N
\!-\!
{\displaystyle \frac{1}{N}}
\right) ,~~
< \!\chi\! >_{\mbox{\scriptsize min}} 
= 
{\displaystyle \frac{1}{N}} .
\EA
\right\}
\label{minimumscalerpot2} 
\eeqa

To find a proper solution for the extended supersymmetric $\sigma$-model,
after rescaling the Goldstone fields ${\cal Q}$ by a mass parameter,
as van Holten et al. did
\cite{NNH.01,vanHolten.85},
we also introduce the $(N+1)$-dimensional matrices 
${\cal R}_f ({\cal Q}_f; \delta {\cal G})$, 
${\cal R}_{fT}({\cal Q}_f; \delta {\cal G})$ and ${\cal X}_f$  
in the following forms:
\beqa
\!\!\!\!\!\!\!\!\!\!\!\!
\left.
\BA{ll}
&
{\cal R}_f ({\cal Q}_f; \delta {\cal G})
\!=\!
{\displaystyle \frac{1}{f}}
\delta {\cal B} 
\!-\! 
\delta {\cal A}^{\mbox{\scriptsize T}}{\cal Q}_f 
\!-\! 
{\cal Q}_f \delta {\cal A}
\!+\! 
f {\cal Q}_f \delta {\cal B}^\dag {\cal Q}_f ,~
{\cal R}_{fT} ({\cal Q}_f; \delta {\cal G})
\!=\!
\!-\!
\delta {\cal A}^{\mbox{\scriptsize T}}
\!+\! 
f {\cal Q}_f \delta {\cal B}^\dag , \!\! \\
\\[-12pt]
&
{\cal X}_f
\!=\!
(1_{N+1} + f^2 {\cal Q}_f {\cal Q}^\dag _f)^{-1}
\!=\! 
{\mathcal X}^\dag ,~
{\cal Q}_f
\!=\! 
\left[ \!\!
\BA{cc}
q & {\displaystyle \frac{1}{f}} r_f \\
- {\displaystyle \frac{1}{f}} r^{\mbox{\scriptsize T}} _f & 0
\EA \!\!
\right] ,~
r_f
\!=\!
{\displaystyle \frac{1}{2 Z^2}}
\left(
x \!+\! f q \bar{x}
\right) ,~
{\displaystyle f 
\!\stackrel{\mathrm{def}}{=}\! 
\frac{1}{m_\sigma }} . \!\!
\EA
\right\}
\label{RRTChif}
\eeqa
Due to the rescaling, 
the Killing potential ${\cal M}_{\sigma }$ is deformed as
\beqa
\left.
\BA{rl}
-i{\cal M}_{f \sigma }
\left(
{\cal Q}_f, \bar{\cal Q}_f;\delta {\cal G}
\right)
&=
-\mbox{tr}
\Delta_f
\left(
{\cal Q}_f, \bar{\cal Q}_f;\delta {\cal G}
\right) ,\\
\\
\Delta_f
\left(
{\cal Q}_f, \bar{\cal Q}_f;\delta {\cal G}
\right)
&\stackrel{\mathrm{def}}{=}
{\cal R}_{fT} ({\cal Q}_f; \delta {\cal G})
-
{\cal R}_f({\cal Q}_f; \delta {\cal G}) f^2 {\cal Q}^\dag _f {\cal X}_f \\
\\
&
=
\left(
f^2 {\cal Q}_f \delta {\cal A} {\cal Q}^\dag _f
- 
\delta {\cal A}^{\mbox{\scriptsize T}}
-
f \delta {\cal B} {\cal Q}^\dag _f
+
f {\cal Q}_f \delta {\cal B}^\dag
\right)
{\cal X}_f ,
\EA
\right\}
\label{formKillingpotMf} 
\eeqa
from which we obtain
a $f$-deformed Killing potential ${\cal M}_{f \sigma }$
\beq
-i{\cal M}_{f \sigma \delta {\cal B}}
=
-f {\cal X}_f {\cal Q}_f ,~~
-i{\cal M}_{f \sigma \delta {\cal B}^\dag }
=
 f {\cal Q}^\dag _f {\cal X}_f ,~~
-i{\cal M}_{f \sigma \delta {\cal A}}
=
1_{N+1} - 2 f^2 {\cal Q}^\dag _f{\cal X}_f {\cal Q}_f .
\label{componentKillingpotMf} 
\eeq
After the same algebraic manipulations, 
the inverse matrix 
${\cal X}_f$
in (\ref{RRTChif}) 
leads to a different form deformed from the previous one
(\ref{inverse1plusQQ})
\beq
{\mathcal X}_f
=
\left[ 
\BA{cc}
{\cal Q}_{f qq^\dag } & {\cal Q}_{f q r} \\
{\cal Q}_{f q r}^{\dag } & {\cal Q}_{f r^\dag r}
\EA 
\right] ,~~
\chi_f
=
(1_N + f^2 q q^\dag )^{-1} 
=
\chi^\dag _f ,
\label{inverse1plusQQf}
\eeq
where each sub-matrix is expressed by the variables $q$ and $r$ 
and the parameters $f$ and $Z$ as
\beq
{\cal Q}_{f qq^\dag }
=
\chi_f - Z^2 \chi_f (r_f r^\dag _f 
- f^2 q\bar{r}_f r^{\mbox{\scriptsize T}}_f q^\dag )
\chi_f ,
\label{inverse1plusQQ1f}
\eeq
\beq
{\cal Q}_{f q \bar{r}}
=
f Z^2 \chi_f q \bar{r}_f ,~~~
{\cal Q}_{f r^\dag r}
=
Z^2 ,
\label{inverse1plusQQsub23f}
\eeq
which are derived in Appendix E.
Substituting 
(\ref{RRTChif}) and (\ref{inverse1plusQQf})
into
(\ref{componentKillingpotMf})
and introducing a $f$-deformed auxiliary function
$
\lambda_f
\!=\!
r_f r_f ^\dag  \!-\! f^2 q\bar{r}_f r_f ^{\mbox{\scriptsize T}}q^\dag
\!=\!
\lambda^\dag _f
$,
we can get the $f$-deformed Killing potential 
${\cal M}_{f \sigma \delta {\cal A}}$
as,
\vspace{-0.1cm}
\beqa
\!\!\!\!\!\!\!\!
\BA{ll}
&~
-i{\cal M}_{f \sigma \delta {\cal A}} = \\
\\
&
\left[ \!\!\!\!\! 
\BA{cc}
\BA{c}
1_N \!-\! 2q^\dag \chi_f q 
\!+\! 
2Z^2 
\left( 
q^\dag \chi_f \lambda_f \chi_f q
+
q^\dag \chi_f q \bar{r}_f r^{\mbox{\scriptsize T}}_f
\right. \\
\left.
+
\bar{r}_f r^{\mbox{\scriptsize T}}_f q^\dag \chi_f q 
\!-\!
{\displaystyle \frac{1}{f^2}}\bar{r}_f r^{\mbox{\scriptsize T}}_f 
\right) 
\EA 
&
\BA{c} 
- 
2 {\displaystyle \frac{1}{f}}q^\dag \chi_f r_f 
\!+\! 
2 {\displaystyle \frac{1}{f}}Z^2 
\left(
q^\dag \chi_f \lambda_f \chi_f r_f
\right. \\
\left.
+
\bar{r}_f r^{\mbox{\scriptsize T}}_f q^\dag \chi_f r_f
\right)  
\EA \\
\\
\BA{c}
- 2 {\displaystyle \frac{1}{f}}r^\dag_f \chi_f q 
\!+\! 
2 {\displaystyle \frac{1}{f}}Z^2 
\left( 
r^\dag_f \chi_f \lambda_f \chi_f q
\right. \!\!\! \\
\left.
+
r^\dag_f \chi_f q \bar{r}_f r^{\mbox{\scriptsize T}}_f 
\right) 
\EA
&
1
\!-\!
2 {\displaystyle \frac{1}{f^2}}r^\dag_f \chi_f r_f 
\!+\! 
2{\displaystyle \frac{1}{f^2}}Z^2 
r^\dag_f \chi_f \lambda_f \chi_f r_f 
\EA \!\!\!
\right] ,
\EA
\label{KillingpotAf} 
\eeqa
in which
each block-matrix satisfy 
the following identities and relations:
\vspace{-0.1cm}
\beq
r^{\mbox{\scriptsize T}}_f q^\dag \chi_f r_f
=
0 ,~~
r^\dag_f \chi_f q \bar{r}_f
=
0 ,~~
r^\dag_f \chi_f r_f
=
\frac{1- Z^2}{Z^2} ,~~
r^\dag_f \chi_f \lambda_f \chi_f r_f
=
\left(
\frac{1- Z^2}{Z^2} 
\right)^2 ,
\label{identitiesf} 
\eeq
\vspace{-0.2cm}
\beq
1
-
2 \frac{1}{f^2} r^\dag_f \chi_f r_f 
+ 
2 \frac{1}{f^2} Z^2 
r^\dag_f \chi_f \lambda_f \chi_f r_f 
=
\frac{1}{f^2} (2Z^2 -1) + 1 - \frac{1}{f^2} ,
\label{relation1f}
\eeq
\vspace{-0.1cm}
\beq
\chi_f \lambda_f \chi_f r_f
=
\frac{1- Z^2}{Z^2} \chi_f r_f ,~~
r^\dag_f \chi_f \lambda_f \chi_f 
=
\frac{1- Z^2}{Z^2} r^\dag_f \chi_f ,~~ 
q^\dag \chi_f q
=
\frac{1}{f^2} (1_N - \bar{\chi }_f) .
\label{relation2f}
\eeq
Using these identities and relations,
we get a more compact form of the $f$-deformed Killing potential 
${\cal M}_{f \sigma \delta {\cal A}}$
as,
\vspace{-0.1cm}
\beq
\!
-i{\cal M}_{f \sigma \delta {\cal A}} 
\!=\! 
\left[ \!\!\!
\BA{cc}
1_N \!-\! 2q^\dag \chi_f q 
\!+\! 
2Z^2 
\left( \!\!
q^\dag \chi_f r_f r^\dag_f \chi_f q
\!-\!
{\displaystyle \frac{1}{f^2}} 
\bar{\chi }_f \bar{r}_f r^{\mbox{\scriptsize T}}_f \bar{\chi }_f \!\!
\right)  & 
- 2 {\displaystyle \frac{1}{f}}Z^2 q^\dag \chi_f r_f \\
\\[-12pt]
- 2 {\displaystyle \frac{1}{f}}Z^2 
r^\dag_f \chi_f q & 
{\displaystyle \frac{1}{f^2}(2Z^2 \!-\! 1) \!+\! 1 \!-\! \frac{1}{f^2}}
\EA \!\!\!
\right] .
\label{KillingpotA2f}
\eeq
Owing to the rescaling,
the $f$-deformed reduced scalar potential is written as follows:
\beqa
\left.
\BA{rl}
&\!\!\!\!
V_{f{\mbox{\scriptsize redSC}}}
=
{\displaystyle \frac{g^2 _{U(1)}}{2(N+1)}}
\left(
\xi - i {\cal M}_{fY}
\right)^2
+
{\displaystyle \frac{g^2 _{SU(N+1)}}{2}}
\mbox{tr}
\left(
- i {\cal M}_{ft}
\right)^2 , \\
\\[-12pt]
&\!\!\!\!
\mbox{tr}
(- i {\cal M}_{ft} )^2
=
\mbox{tr}
(- i {\cal M}_{f \sigma \delta {\cal A}})^2
-
{\displaystyle \frac{1}{N+1}}
(- i {\cal M}_{fY})^2 ,~~
- i {\cal M}_{fY}
=
\mbox{tr}
\left(
- i {\cal M}_{f \sigma \delta {\cal A}} 
\right),
\EA
\right\} 
\label{specialscalerpotf} 
\eeqa
in which each $f$-deformed Killing potential is calculated 
straight forwardly
in the following forms:
\beqa
\BA{rl}
\mbox{tr}
\left(
-i{\cal M}_{f \sigma \delta {\cal A}}
\right)
=
&\!\!\!\!
\left( \!
1 - 2 {\displaystyle \frac{1}{f^2}} \!
\right) \!
N 
+ 
2 {\displaystyle \frac{1}{f^2}} \mbox{{\bf tr}}(\chi_f)
+
2 {\displaystyle \frac{1}{f^2}} Z^2 
\mbox{{\bf tr}}(\chi_f r_f r^\dag_f)
-
4 {\displaystyle \frac{1}{f^2}} Z^2 
\mbox{{\bf tr}}(\chi_f r_f r^\dag_f \chi_f) \\
\\
&\!\!\!\!
+
{\displaystyle \frac{1}{f^2}} (2 Z^2 - 1) 
+ 
1 - {\displaystyle \frac{1}{f^2}} , 
\EA
\label{trM2f1}
\eeqa
\beqa
\BA{rl}
&\!\!\!\!
\mbox{tr}
\left(
-i{\cal M}_{f \sigma \delta {\cal A}}
\right)^2
=
N - 4 {\displaystyle \frac{1}{f^2}} (1 - {\displaystyle \frac{1}{f^2}}) N
- 
4 {\displaystyle \frac{1}{f^4}} \mbox{{\bf tr}}(\chi_f )
+
4 {\displaystyle \frac{1}{f^4}} \mbox{{\bf tr}}(\chi_f \chi_f) \\
\\
&\!\!\!\!
+
4 {\displaystyle \frac{1}{f^2}} (1 - {\displaystyle \frac{1}{f^2}}) Z^2
\mbox{{\bf tr}}(\chi_f r_f r^\dag_f) 
- 
4 {\displaystyle \frac{1}{f^2}} (1 - {\displaystyle \frac{1}{f^2}}) Z^2
\mbox{{\bf tr}}(\chi_f r_f r^\dag_f \chi_f) \\
\\
&\!\!\!\!
+
12 {\displaystyle \frac{1}{f^4}} Z^2 
\mbox{{\bf tr}}(\chi_f r_f r^\dag_f \chi_f) 
-
16 {\displaystyle \frac{1}{f^4}} Z^2 
\mbox{{\bf tr}}(\chi_f \chi_f r_f r^\dag_f \chi_f) \\
\\
&\!\!\!\!
-
4 {\displaystyle \frac{1}{f^4}} Z^4 r^\dag_f \chi_f \chi_f r_f \cdot 
\mbox{{\bf tr}}(\chi_f r_f r^\dag_f )
+
8 {\displaystyle \frac{1}{f^4}} Z^4 r^\dag_f \chi_f \chi_f r_f \cdot 
\mbox{{\bf tr}}(\chi_f r_f r^\dag_f \chi_f) \\
\\
&\!\!\!\!
+
{\displaystyle \frac{1}{f^4}} 
+
2 {\displaystyle \frac{1}{f^2}}
(1 - {\displaystyle \frac{1}{f^2}}) (2 Z^2 - 1)
+
(1 - {\displaystyle \frac{1}{f^2}})^2 
- 
4 {\displaystyle \frac{1}{f^4}} Z^4 r^\dag_f \chi_f \chi_f r_f .
\EA
\label{trM2f2}
\eeqa
The identity below is also derived
\beqa
\BA{rl}
r^\dag_f \chi_f r_f
=
&\!\!\!\!
{\displaystyle 
\frac{1}{4Z^4}
(x^\dag + f x^{\mbox{\scriptsize T}} q^\dag) \chi_f (x + f q \bar{x})
=
\frac{1}{4Z^4}
(x^\dag \chi_f x + x^{\mbox{\scriptsize T}} q^\dag \chi_f q \bar{x})
} \\
\\
=
&\!\!\!\!
{\displaystyle 
\frac{1}{4Z^4}
x^{\mbox{\scriptsize T}} \bar{x}
=
\frac{1 - z^2}{4Z^4}
=
\frac{1 - Z^2}{Z^2}
} ,
\EA
\label{rdagchirf}
\eeqa
and approximate formulas for the quantities
$r^\dag_f  \chi_f \chi_f r_f$ and $\mbox{{\bf tr}}(r_f r^\dag_f )$
can be calculated  
as
\beqa
\left.
\BA{rl}
r^\dag_f  \chi_f \chi_f r_f
=
&\!\!\!\!
{\displaystyle
\frac{1}{4Z^4}
(x^\dag + f x^{\mbox{\scriptsize T}} q^\dag) 
\chi_f \chi_f (x + f q \bar{x})
=
\frac{1}{4Z^4}
x^\dag \chi_f x 
} \\
\\
\approx
&\!\!\!\!
{\displaystyle \frac{1}{4Z^4}}
\left\{
\frac{1}{N}
\left[
N + f^2 \mbox{{\bf tr}}(q^\dag q)
\right]
\right\}^{-1}
x^\dag x
=
{\displaystyle \frac{1 - Z^2}{Z^2}}
< \!\chi_f\! >, \\
\\
< \!\chi_f\! >
\stackrel{\mathrm{def}}{=}
&\!\!\!\!
\left\{
{\displaystyle \frac{1}{N}}
\left[
N + f^2 \mbox{{\bf tr}}(q^\dag q)
\right]
\right\}^{-1} ,
\EA
\right\}
\label{rdagchichirf}
\eeqa
\beqa
\BA{rl}
\mbox{{\bf tr}}(r_f r^\dag_f )
=
&\!\!\!\!
r^\dag_f r_f
=
{\displaystyle
\frac{1}{4Z^4}
(x^\dag + f x^{\mbox{\scriptsize T}} q^\dag) (x + f q \bar{x})
=
\frac{1}{4Z^4}
x^\dag \chi_f ^{-1} x 
} \\
\\
\approx
&\!\!\!\!
{\displaystyle
\frac{1 - Z^2}{Z^2}
\frac{1}{< \!\chi_f\! >}
} 
\stackrel{\mathrm{def}}{=}
< \!r_f r^\dag_f\! > .
\EA
\label{trrrdagf}
\eeqa
In 
(\ref{trM2f1}) and (\ref{trM2f2}),
approximating
$\mbox{{\bf tr}}(\chi_f )$,
$\mbox{{\bf tr}}(\chi_f r_f r^\dag_f )$, etc.
by
$< \!\!\chi_f\!\! >$,
$< \!\!\chi_f\!\! > \! \mbox{{\bf tr}}(r_f r^\dag_f )$, etc.,
respectively,
and using
(\ref{rdagchichirf}) and (\ref{trrrdagf}),
$
\mbox{tr}
\left(
-i{\cal M}_{f \sigma \delta {\cal A}}
\right)
$ 
and 
$
\mbox{tr}
\left(
-i{\cal M}_{f \sigma \delta {\cal A}}
\right)^2
$
are computed as
\beqa
\!\!\!\!\!\!
\left.
\BA{rl}
\mbox{tr}
\left(
-i{\cal M}_{f \sigma \delta {\cal A}}
\right)
=
&\!\!\!\!
1 
+
\left( \!
1 - 2 {\displaystyle \frac{1}{f^2}} \!
\right) \!
N 
+ 
2 {\displaystyle \frac{1}{f^2}} (2Z^2 \!-\! 1) \!
< \!\chi_f\! > , \\
\\
\mbox{tr}
\left(
-i{\cal M}_{f \sigma \delta {\cal A}}
\right)^2
=
&\!\!\!\!
1 + N - 4 {\displaystyle \frac{1}{f^2}} 
\left( \!
1 - {\displaystyle \frac{1}{f^2}} \!
\right) \! N \\
\\
-
&\!\!\!\!
4 {\displaystyle \frac{1}{f^2}}
\left\{
{\displaystyle \frac{1}{f^2}}(2 Z^2 - 1)
-
\left( \!
1 - {\displaystyle \frac{1}{f^2}} \!
\right) \!
Z^2 \!
\right\} \!
< \!\chi_f\! >
+ 
4 {\displaystyle \frac{1}{f^4}}(2 Z^4 - 1) \! < \!\chi_f\! >^2 .
\EA \!\!
\right\}
\label{approxKillingpotf}
\eeqa
Substituting
(\ref{approxKillingpotf})
into
(\ref{specialscalerpotf}),
we obtain the $f$-deformed reduced scalar potential
as
\beqa
\!\!\!\!\!\!
\BA{rl}
&\!\!\!\!
V_{f{\mbox{\scriptsize redSC}}}
=
{\displaystyle
\frac{g^2 _{U(1)}}{2(N \!+\! 1)}
}
\left[
\xi + 1 
+ 
\left(1 - 2 {\displaystyle \frac{1}{f^2}}\right) 
N 
+ 
2
{\displaystyle \frac{1}{f^2}}(2 Z^2 - 1)  
< \!\chi_f\! >
\right]^2 \\
\\[-6pt]
&\!\!\!\!
+
{\displaystyle
2 
\frac{g^2 _{SU(N + 1)}}{N + 1}
}
{\displaystyle \frac{1}{f^2}}
\left[
{\displaystyle \frac{1}{f^2}} N 
- 
\left\{
\left(
1 - {\displaystyle \frac{1}{f^2}}
\right) N
+
\left(
1 + 3 {\displaystyle \frac{1}{f^2}}
\right)
\right\}
Z^2 \! < \!\chi_f\! > 
\right. \\
\\[-6pt]
&\!\!\!\!
\left.
+
\left\{
\left(
1 - {\displaystyle \frac{1}{f^2}}
\right) N
+
\left(
1 + {\displaystyle \frac{1}{f^2}}
\right)
\right\}
< \!\chi_f\! >
+
{\displaystyle \frac{1}{f^2}}
\left\{
2(N - 1) Z^4 + 4 Z^2 - (N + 2)
\right\} \!
< \!\chi_f\! >^2
\right] ,
\EA
\label{specialscalerpot2f} 
\eeqa
and, from the variation of this
with respect to $Z$ and $< \!\!\chi_f\!\! >$,
we get the following relations:
\beqa
\BA{rl}
&\!\!\!\!
g^2 _{U(1)}
\left\{
\xi + 1 
+ 
\left(1 - 2 {\displaystyle \frac{1}{f^2}}\right) 
N 
+ 
2
{\displaystyle \frac{1}{f^2}}(2 Z^2 - 1)  
< \!\chi_f\! >
\right\} \\
\\[-6pt]
&\!\!\!\!
-
2 
g^2 _{SU(N \!+\! 1)}
\left[
{\displaystyle \frac{1}{4}}
\left\{

\left(
1 - {\displaystyle \frac{1}{f^2}}
\right) N
+
\left(
1 + 3 {\displaystyle \frac{1}{f^2}}
\right)
\right\}
- {\displaystyle \frac{1}{f^2}}
\left\{
(N - 1) Z^2 + 1
\right\}
< \!\chi_f\! > 
\right]
= 0 ,
\EA
\label{VvariZf} 
\eeqa
\beqa
\BA{rl}
&\!\!\!\!
g^2 _{U(1)}
\left[
\xi + 1 
+ 
\left(1 - 2 {\displaystyle \frac{1}{f^2}}\right) 
N 
+ 
2
{\displaystyle \frac{1}{f^2}}(2 Z^2 - 1)  
< \!\chi_f\! >
\right]
(2 Z^2 - 1)  \\
\\[-6pt]
&\!\!\!\!
-
2 
g^2 _{SU(N \!+\! 1)}
\left[
{\displaystyle \frac{1}{2}}
\left\{
\left(
1 - {\displaystyle \frac{1}{f^2}}
\right) N
+
\left(
1 + 3 {\displaystyle \frac{1}{f^2}}
\right)
\right\}
Z^2
-
{\displaystyle \frac{1}{2}}
\left(
1 - {\displaystyle \frac{1}{f^2}}
\right) N
-
{\displaystyle \frac{1}{2}}
\left(
1 + {\displaystyle \frac{1}{f^2}}
\right)
\right.\\
\\[-6pt]
&\!\!\!\!
\left.
~~~~~~~~~~~~~~~~~~~~~~~~~
- {\displaystyle \frac{1}{f^2}}
\left\{
2 (N - 1) Z^4 + 4 Z^2 - (N + 2)
\right\}
< \!\chi_f\! > 
\right]
= 0 .
\EA
\label{Vvarichif}
\eeqa
Multiplying by
$(2 Z^2 \!-\! 1)$
for
(\ref{VvariZf}) 
and using 
(\ref{Vvarichif}),
we have a $g^2$-independent relation
\beqa
\BA{rl}
&
\left[
{\displaystyle \frac{1}{4}}
\left\{

\left(
1 - {\displaystyle \frac{1}{f^2}}
\right) N
+
1 + 3 {\displaystyle \frac{1}{f^2}}
\right\}
- {\displaystyle \frac{1}{f^2}}
\left\{
(N - 1) Z^2 + 1
\right\}
< \!\chi_f\! > 
\right]
(2Z^2 \!-\! 1) \\
\\[-6pt]
&
-
\left[
{\displaystyle \frac{1}{2}}
\left\{
\left(
1 - {\displaystyle \frac{1}{f^2}}
\right) N
+
\left(
1 + 3 {\displaystyle \frac{1}{f^2}}
\right)
\right\}
Z^2
-
{\displaystyle \frac{1}{2}}
\left(
1 - {\displaystyle \frac{1}{f^2}}
\right) N
-
{\displaystyle \frac{1}{2}}
\left(
1 + {\displaystyle \frac{1}{f^2}}
\right)
\right.\\
\\[-6pt]
&\!\!\!\!
\left.
~~~~~~~~~~~~~~~~~~~~~~~~~
- {\displaystyle \frac{1}{f^2}}
\left\{
2 (N - 1) Z^4 + 4 Z^2 - (N + 2)
\right\}
< \!\chi_f\! > 
\right]
= 0 .
\EA
\label{eqdetermineZ2f} 
\eeqa
This relation reads
\beq
(N + 1)
\left\{
4 (Z^2 - 1) < \!\chi_f\! > 
-
\left( 1 - f^2 \right)
\right\}
= 0 ,
\label{relationforZ2chif}
\eeq
through which
due to $0 \! \le \!Z^2\! \le \! 1$ and $< \!\!\chi_f\!\! > \! > \! 0$,
the rescaling parameter $f$ is shown to satisfy $f^2 \!\! \ge \!\! 1$.
From (\ref{relationforZ2chif}), 
since $< \!\!\chi_f\!\! > \neq 0$,
we can finally reach our ultimate goal of proper solutions 
for $Z^2$ and $< \!\!\chi_f\!\! >$ as
\beqa
\!\!\!\!\!\!\!\!\!\!\!\!
\left.
\BA{rl}
&Z^2 
= 
1
+
{\displaystyle \frac{1}{2}} 
\left( 1 - f^2 \right)
{\displaystyle 
\frac{g^2 _{U(1)} \!+\! N g^2 _{SU(N \!+\! 1)}}
{g^2 _{U(1)}
\left\{ 
(2 \!-\! f^2) N \!-\! 1
\right\} 
\!-\! 
g^2 _{SU(N \!+\! 1)}
\left\{
(1 \!-\! f^2) N \!-\! 2
\right\}
\!-\!
g^2 _{U(1)} f^2 \xi }
} , \\
\\[-8pt]
&< \!\!\chi_f\!\! > 
=\!
{\displaystyle  
\frac{1}{2}
\frac{1}{g^2 _{U(1)} \!+\! N g^2 _{SU(N \!+\! 1)}}
} \!\!
\left[
g^2 _{U(1)} \!
\left\{ 
(2 \!-\! f^2) N \!-\! 1
\right\} 
\!-\! 
g^2 _{SU(N \!+\! 1)} \!
\left\{
(1 \!-\! f^2) N \!-\! 2
\right\}
\!-\!
g^2 _{U(1)} f^2 \xi 
\right] \! .
\EA \!\!\!
\right\}
\label{solutionforZ2chif}
\eeqa
This is just the solution for the $\frac{SO(2N+2)}{U(N+1)}$
supersymmetric $\sigma$-model.
The solution 
(\ref{solutionforZ2chif}),
if $f^2 \!=\! 1$,
reduces to a simple solution
(\ref{simplesolutionforchi}).


\newpage

\setcounter{equation}{0}
\renewcommand{\theequation}{\arabic{section}.\arabic{equation}}

\section{Discussions and concluding remarks}

~~~In order to find a proper solution for 
the extended $\frac{SO(2N+2)}{U(N+1)}$ supersymmetric $\sigma$-model,
the minimization of the $f$-deformed reduced scalar potential 
has been made
after rescaling Goldstone fields by a mass parameter.
Then the proper solutions for
$Z^2$ and $< \!\!\chi_f\!\! >$
(\ref{solutionforZ2chif})
have been produced.
In the course of producing such solutions, 
the Fayet-Ilipoulos term has made a crucial role.
Substituting
(\ref{solutionforZ2chif})
into
(\ref{specialscalerpot2f}),
the minimization of 
the $f$-deformed reduced scalar potential
with respect to 
the Fayet-Ilipoulos parameter $\xi$
is realized
as follows:
\beqa
\!\!\!\!
\left.
\BA{rl}
&\!\!\!\!
V_{f{\mbox{\scriptsize redSC}}}
=
{\displaystyle \frac{1}{2}}
{\displaystyle \frac{N}{N \!+\! 1}}
{\displaystyle
\frac{g^2 _{U(1)}g^2 _{SU(N + 1)}}{g^2 _{U(1)} \!+\! N g^2 _{SU(N + 1)}}
}
\left[
\xi 
\!+\!
{\displaystyle \frac{1}{N}} 
\left\{
\left(
1 \!-\! 2 {\displaystyle \frac{1}{f^2}}
\right)
N^2  
\!+\!
N
\!+\!
2 {\displaystyle \frac{1}{f^2}} 
\right\}
\right]^2 
\!+\!
V_{f{\mbox{\scriptsize redSC}}}^{\mbox{\scriptsize min}} ,
\\
\\[-10pt]
&\!\!\!\!
~~~
\xi_{\mbox{\scriptsize min}} 
=
-
\left(
1 \!-\! 2 {\displaystyle \frac{1}{f^2}}
\right)
N
\!-\!
1
\!-\! 
2 {\displaystyle \frac{1}{f^2}}
{\displaystyle \frac{1}{N}}   ,
\\
\\[-10pt]
&\!\!\!\!
V_{f{\mbox{\scriptsize redSC}}}^{\mbox{\scriptsize min}}
\!=\!
{\displaystyle
2 
\frac{g^2 _{SU(N + 1)}}{N + 1}
}
\left[
\left\{
{\displaystyle \frac{1}{f^4}}
\!+\! 
{\displaystyle \frac{1}{8}}
\left(
1 \!-\! {\displaystyle \frac{1}{f^2}}
\right) ^2 
\right\} N
\!+\!
{\displaystyle \frac{1}{8}}
\left(
1 \!-\! {\displaystyle \frac{1}{f^2}}
\right) ^2
\!-\!
{\displaystyle \frac{1}{f^4}}
{\displaystyle \frac{1}{N}}
\right] .
\EA
\right\}
\label{minimumscalerpot2f} 
\eeqa
Thus we get the minimized $f$-deformed reduced scalar potential
$V_{f{\mbox{\scriptsize redSC}}}^{\mbox{\scriptsize min}}$
if we choose the Fayet-Ilipoulos parameter $\xi$ to be
$\xi_{\mbox{\scriptsize min}}$.
Putting this
$\xi_{\mbox{\scriptsize min}}$
into 
(\ref{solutionforZ2chif}),
we have
\beqa
\!\!\!\!\!\!\!\!
\left.
\BA{rl}
&Z_{\mbox{\scriptsize min}}^2 
= 
{\displaystyle \frac{1}{2}} 
+
{\displaystyle \frac{1}{2}}
{\displaystyle \frac{1}{N}}  
{\displaystyle 
\frac{1}
{
{\displaystyle \frac{1}{2}} 
(f^2 - 1)
+
{\displaystyle \frac{1}{N}} 
}
} , \\
\\[-12pt]
&< \!\!\chi_f\!\! >_{\mbox{\scriptsize min}} 
=
{\displaystyle \frac{1}{2}} 
(f^2 - 1)
+
{\displaystyle \frac{1}{N}} ,~~f^2 \ge 1 .
\EA
\right\}
\label{solutionforZ2chifmini}
\eeqa 
Equations
(\ref{minimumscalerpot2f}) and (\ref{solutionforZ2chifmini}),
if $f^2 \!=\! 1$,
reduce to
(\ref{minimumscalerpot2}).
 
In this paper,
we have given an extended supersymmetric $\sigma$-model
on the K\"{a}hler coset space 
$\frac{G}{H} \!=\! \frac{SO(2N \!+\! 2)}{U(N \!+\! 1)}$,
basing on the $SO(2N \!+\! 1)$ Lie algebra of the fermion operators.
Embedding the $SO(2N \!+\! 1)$ group into an $SO(2N \!+\! 2)$ group and 
using the $\frac{SO(2N \!+\! 2)}{U(N \!+\! 1)}$ coset variables  
\cite{Fuk.77},
we have investigated a new aspect of 
the extended supersymmetric $\sigma$-model 
which has never been seen
in the usual supersymmetric $\sigma$-model on the K\"{a}hler coset space 
$\frac{SO(2N)}{U(N)}$
given by van Holten et al.
\cite{NNH.01}.
We have constructed a Killing potential, 
the extension of 
the Killing potential in the $\frac{SO(2N)}{U(N)}$ coset space 
to that in the $\frac{SO(2N \!+\! 2)}{U(N \!+\! 1)}$ coset space.
To our great surprise,
the Killing potential is equivalent with the generalized density matrix
which is a useful tool to study fermion many-body problems.
Its diagonal-block part is related to a reduced scalar potential 
with the Fayet-Ilipoulos term.
The reduced 
and
the $f$-deformed reduced scalar potentials
have been optimized in order to see the behaviour of
the vacuum expectation value of the $\sigma$-model fields.
We have got, if $f^2 \!=\! 1$, a simple solution 
$Z^2 \!=\! 1$
corresponding to the $\frac{SO(2N)}{U(N)}$ 
supersymmetric $\sigma$-model
and
the proper solutions for $Z^2$ and $< \!\!\chi_f\!\! >$.
The Fayet-Ilipoulos term has made an important role to get
such solutions.

Finally,
we have given
bosonization 
of the $SO(2N \!+\! 2)$ Lie operators,
vacuum functions and differential forms for their bosons expressed 
in terms of the $\frac{SO(2N+2)}{U(N+1)}$ coset variables,
a $U(1)$ phase
and the corresponding K\"{a}hler potential.
This provides a powerful tool for describing the Goldstone bosons
but accompanying fermionic modes
in the present model.
The effectiveness of $\frac{SO(2N+2)}{U(N+1)}$ K\"{a}hler manifold 
is expected to open a new field for exploration of 
low-energy elementary particle physics
by the supersymmetric $\sigma$-model.


\newpage

\vskip0.5cm
\begin{center}
{\bf Acknowledgements}
\end{center}
~~~~One of the authors (S. N.) would like to
express his sincere thanks to
Professor Alex H. Blin for kind and
warm hospitality extended to
him at the Centro de F\'\i sica Te\' orica,
Universidade de Coimbra, Portugal.
This work was supported by the Portuguese Project
POCTI/FIS/451/94.
The authors thank the Yukawa Institute for Theoretical Physics
at Kyoto University. Discussions during the YITP workshop
YITP-W-07-05 on ``String Theory and Quantum Field Theory''
were useful to complete this work.


\newpage

\leftline{\large{\bf Appendix}}
\appendix


\def\thesection{\Alph{section}}
\setcounter{equation}{0}
\renewcommand{\theequation}{\Alph{section}.\arabic{equation}}
\section{Bosonization of SO(2N+2) Lie operators}


~~~Consider 
a fermion state vector $\ket \Psi$ 
corresponding to a function 
$\Psi ({\cal G})$ in ${\cal G} \in SO(2N+2)$:  
\beqa
\BA{l}
\ket \Psi
=
\int U ({\cal G}) \ket 0 \bra 0 U^\dag ({\cal G}) \ket \Psi d{\cal G}
=
\int U ({\cal G}) \ket 0 \Psi ({\cal G}) d{\cal G} .
\EA
\label{statePsi}
\eeqa
The ${\cal G}$ is given by (\ref{calAcalB}) and (\ref{calG}) and
the $d{\cal G}$ is an invariant group integration.  
When an infinitesimal operator 
$\mathbb{I}_{\cal G} \!+\! \delta \widehat{{\cal G}}$ and
a corresponding infinitesimal unitary operator 
$U (1_{2N+2} \!+\! \delta {\cal G})$
is operated on $\ket \Psi$,
using
$U^{-1}(1_{2N+2} \!+\! \delta {\cal G}) 
\!=\! 
U (1_{2N+2} \!-\! \delta {\cal G})$, 
it transforms $\ket \Psi$ as
\beqa
\!\!\!\!\!\!
\BA{ll}
&U (1_{2N+2} - \delta {\cal G})  \ket \Psi
=
(\mathbb{I}_{\cal G} - \delta \widehat{{\cal G}}) \ket \Psi
=
\int U ({\cal G}) \ket 0 \bra 0 
U^\dag ((1_{2N+2} + \delta {\cal G}){\cal G}) \ket \Psi d{\cal G} \\
\\[-8pt]
&=
\int U ({\cal G}) \ket 0 \Psi ((1_{2N+2} 
+ 
\delta {\cal G}){\cal G})  d{\cal G} 
=
\int U ({\cal G}) \ket 0 (1_{2N+2} + \delta \mbox{\boldmath ${\cal G}$})  
\Psi ({\cal G}) d{\cal G},
\EA
\label{infinitrans}
\eeqa 
\beqa
\!\!\!\!\!\!
\left.
\BA{ll}
&
1_{2N+2} + \delta {\cal G}
=
\left[ 
\BA{cc} 
1_{N+1} + \delta {\cal A} & \delta \bar{\cal B}\\
\delta {\cal B} & 1_{N+1} + \delta \bar{\cal A} \\ 
\EA 
\right] ,~
\delta {\cal A}^\dag = - \delta {\cal A},~
\mbox{tr}\delta {\cal A} = 0 ,~
\delta {\cal B} = - \delta {\cal B}^{\mbox{\scriptsize T}} , \\
\\[-8pt]
&\delta \widehat{{\cal G}}
=
\delta {\cal A}^p_{~q} E^q_{~p}
+
{\displaystyle \frac{1}{2}} 
\left(
\delta {\cal B}_{pq} E^{qp} + \delta \bar{\cal B}_{pq} E_{qp}
\right) ,~
\delta \mbox{\boldmath ${\cal G}$}
=
\delta {\cal A}^p_{~q} \mbox{\boldmath ${\cal E}^q_{~p}$}
+
{\displaystyle \frac{1}{2}} 
\left(
\delta {\cal B}_{pq} \mbox{\boldmath ${\cal E}^{qp}$}
+ 
\delta \bar{\cal B}_{pq} \mbox{\boldmath ${\cal E}_{qp}$}
\right) . \!\!
\EA
\right\}
\label{infinitesimalop}
\eeqa
Equation
(\ref{infinitrans})
shows that
the operation of $\mathbb{I}_{\cal G} \!-\! \delta \widehat{{\cal G}}$ 
on the $\ket \Psi$ in the fermion space 
corresponds to the left multiplication by $1_{2N+2} \!+\! \delta {\cal G}$
for the variable of the ${\cal G}$ of the function $\Psi ({\cal G})$.
For a small parameter $\epsilon$,
we obtain a representation on the $\Psi ({\cal G})$ as
\beq
\rho (e^{\epsilon \delta {\cal G}}) \Psi ({\cal G})
=
\Psi (e^{\epsilon \delta {\cal G}} {\cal G})
=
\Psi ({\cal G} + \epsilon \delta {\cal G} {\cal G})
=
\Psi ({\cal G} + d{\cal G}) ,
\label{reponPsi}
\eeq
which leads us to a relation 
$d{\cal G} 
= 
\epsilon \delta {\cal G} {\cal G}$.
From this,
we express it explicitly as,
\beqa
\left.
\BA{ll}
&
d{\cal G}
=
\left[ 
\BA{cc} 
d{\cal A} & d\bar{\cal B} \\
d{\cal B} & d\bar{\cal A} \\ 
\EA 
\right]
=
\epsilon
\left[ 
\BA{cc} 
\delta {\cal A}{\cal A} + \delta \bar{\cal B} {\cal B}  & 
\delta {\cal A}\bar{\cal B} + \delta \bar{\cal B}\bar{\cal A} \\
\delta {\cal B}{\cal A} + \delta \bar{\cal A}{\cal B} & 
\delta \bar{\cal A}\bar{\cal A} + \delta {\cal B}\bar{\cal B} \\ 
\EA 
\right] ,\\
\\
&
d{\cal A}
=
\epsilon
{\displaystyle \frac{\partial {\cal A}}{\partial \epsilon }}
=
\epsilon
(\delta {\cal A}{\cal A} + \delta \bar{\cal B}{\cal B}) ,~~
d{\cal B}
=
\epsilon
{\displaystyle \frac{\partial {\cal B}}{\partial \epsilon }}
=
\epsilon
(\delta {\cal B}{\cal A} + \delta \bar{\cal A}{\cal B}) .
\EA
\right\}
\label{dGdAdB} 
\eeqa
A differential representation of $\rho (\delta {\cal G})$,
$d\rho (\delta {\cal G})$,
is given as
\beq
d\rho (\delta {\cal G}) \Psi ({\cal G})
=
\left[
\frac{\partial {\cal A}^p_{~q}}{\partial \epsilon }
\frac{\partial }{\partial {\cal A}^p_{~q} }
+
\frac{\partial {\cal B}_{pq}}{\partial \epsilon }
\frac{\partial }{\partial {\cal B}_{pq} }
+
\frac{\partial \bar{\cal A}^{p}_{~q}}{\partial \epsilon }
\frac{\partial }{\partial \bar{\cal A}^{p}_{~q} }
+
\frac{\partial \bar{\cal B}_{pq}}{\partial \epsilon }
\frac{\partial }{\partial \bar{\cal B}_{pq}}
\right]
\Psi ({\cal G}) .
\label{diffrep} 
\eeq
Substituting
(\ref{dGdAdB})
into
(\ref{diffrep}),
we can get explicit forms of the differential representation
\beq
d\rho (\delta {\cal G}) \Psi ({\cal G})
=
\delta \mbox{\boldmath ${\cal G}$}
\Psi ({\cal G}),
\eeq
where each operator in $\delta \mbox{\boldmath ${\cal G}$}$
is expressed in a differential form as
\beqa
\left.
\BA{ll}
&
\mbox{\boldmath ${\cal E}^p_{~q}$}
=
\bar{\cal B}_{pr}
{\displaystyle \frac{\partial }{\partial \bar{\cal B}_{qr}}}
-
{\cal B}_{qr}
{\displaystyle \frac{\partial }{\partial {\cal B}_{pr}}}
-
\bar{\cal A}^{q}_{~r}
{\displaystyle \frac{\partial }{\partial \bar{\cal A}^{p}_{~r}}}
+
{\cal A}^p_{~r}
{\displaystyle \frac{\partial }{\partial {\cal A}^q_{~r}}} 
=
\mbox{\boldmath ${\cal E}^{q \dag }_{~p}$} ,\\
\\[-10pt]
&
\mbox{\boldmath ${\cal E}_{pq}$}
=
\bar{\cal A}^{p}_{~r}
{\displaystyle \frac{\partial }{\partial \bar{\cal B}_{qr}}}
-
{\cal B}_{qr}
{\displaystyle \frac{\partial }{\partial {\cal A}^p_{~r}}}
-
\bar{\cal A}^{q}_{~r}
{\displaystyle \frac{\partial }{\partial \bar{\cal B}_{pr}}}
+
{\cal B}_{pr}
{\displaystyle \frac{\partial }{\partial {\cal A}^q_{~r}}} 
=
\mbox{\boldmath ${\cal E}^{qp \dag }$} ,\\
\\[-8pt]
&
\mbox{\boldmath ${\cal E}^{p \dag }_{~q}$}
=
-
\mbox{\boldmath $\bar{{\cal E}}^{p }_{~q}$},~~~~~~~
\mbox{\boldmath ${\cal E}_{pq }^\dag $}
=
-
\mbox{\boldmath $\bar{{\cal E}}_{pq }$},~~~~~~~
\mbox{\boldmath ${\cal E}_{pq}$}
=
-
\mbox{\boldmath ${\cal E}_{qp}$} .
\EA
\right\}
\label{diffops}
\eeqa
We define the boson operators 
$\mbox{\boldmath ${\cal A}^p_{~q}$}$,
$\mbox{\boldmath $\bar{\cal A}^{p}_{~q}$}$, etc.,
from every variable
${\cal A}^p_{~q}$,
$\bar{\cal A}^{p}_{~q}$, etc.,
as
\beqa
\left.
\BA{ll}
&
\mbox{\boldmath ${\cal A}$}
\stackrel{\mathrm{def}}{=}
{\displaystyle
\frac{1}{\sqrt{2}}
\left(
{\cal A} + \frac{\partial }{\partial \bar{\cal A} }
\right) 
} ,~~
\mbox{\boldmath ${\cal A}^\dag$}
\stackrel{\mathrm{def}}{=}
{\displaystyle
\frac{1}{\sqrt{2}}
\left(
\bar{\cal A} - \frac{\partial }{\partial {\cal A}}
\right) 
} ,\\
\\
&
\mbox{\boldmath $\bar{\cal A}$}
\stackrel{\mathrm{def}}{=}
{\displaystyle
\frac{1}{\sqrt{2}}
\left(
\bar{\cal A} + \frac{\partial }{\partial {\cal A}}
\right) 
} ,~~
\mbox{\boldmath ${\cal A}^{\mbox{\scriptsize T}}$}
\stackrel{\mathrm{def}}{=}
{\displaystyle
\frac{1}{\sqrt{2}}
\left(
{\cal A} - \frac{\partial }{\partial \bar{\cal A} }
\right) 
} , \\
\\
&
[\mbox{\boldmath ${\cal A}$},~\mbox{\boldmath ${\cal A}^\dag$}]
=
1 ,~~
[\mbox{\boldmath $\bar{\cal A}$},~
\mbox{\boldmath ${\cal A}^{\mbox{\scriptsize T}}$}]
=1 ,\\
\\
&
[\mbox{\boldmath ${\cal A}$},~\mbox{\boldmath $\bar{\cal A}$}]
=
[\mbox{\boldmath ${\cal A}$},~
\mbox{\boldmath ${\cal A}^{\mbox{\scriptsize T}}$}]
=
0 ,~~
[\mbox{\boldmath ${\cal A}$}^\dag ,~\mbox{\boldmath $\bar{\cal A}$}]
=
[\mbox{\boldmath ${\cal A}$}^\dag ,~
\mbox{\boldmath ${\cal A}^{\mbox{\scriptsize T}}$}]
=
0 ,
\EA
\right\}
\label{boseops}
\eeqa
where ${\cal A}$ is a complex variable.
Similar definitions hold for ${\cal B}$
in order to define the boson operators 
$\mbox{\boldmath ${\cal B}_{pq}$}$,
$\mbox{\boldmath $\bar{\cal B}_{pq}$}$, etc.
By noting the relations
\beq
{\displaystyle
\bar{\cal A} \frac{\partial }{\partial \bar{\cal A} }
-
{\cal A} \frac{\partial }{\partial {\cal A}}
=
\mbox{\boldmath ${\cal A}^\dag$}
\mbox{\boldmath ${\cal A}$}
-
\mbox{\boldmath ${\cal A}^{\mbox{\scriptsize T}}$}
\mbox{\boldmath $\bar{\cal A}$} 
},~~~~
{\displaystyle
\bar{\cal A} \frac{\partial }{\partial \bar{\cal B} }
-
{\cal B} \frac{\partial }{\partial {\cal A}}
=
\mbox{\boldmath ${\cal A}^\dag$}
\mbox{\boldmath ${\cal B}$}
-
\mbox{\boldmath ${\cal B}^{\mbox{\scriptsize T}}$}
\mbox{\boldmath $\bar{\cal A}$} 
},
\eeq
the differential operators
(\ref{diffops})
can be converted into a boson operator representation
\beqa
\left.
\BA{ll}
&
\mbox{\boldmath ${\cal E}^p_{~q}$}
=
\mbox{\boldmath ${\cal B}^\dag _{pr}$}\mbox{\boldmath ${\cal B}_{qr}$}
-
\mbox{\boldmath ${\cal B}^{\mbox{\scriptsize T}}_{qr}$}
\mbox{\boldmath $\bar{\cal B}_{pr}$}
-
\mbox{\boldmath ${\cal A}^{q \dag }_{~r}$}\mbox{\boldmath ${\cal A}^p_{~r}$}
+
\mbox{\boldmath ${\cal A}^{p \mbox{\scriptsize T}}_{~r}$}
\mbox{\boldmath $\bar{\cal A}^{q}_{~r}$} 
=
\mbox{\boldmath ${\cal B}^\dag _{p \tilde{r}}$}
\mbox{\boldmath ${\cal B}_{q \tilde{r}}$}
-
\mbox{\boldmath ${\cal A}^{q \dag }_{~\tilde{r}}$}
\mbox{\boldmath ${\cal A}^p_{~\tilde{r}}$} ,\\
\\
&
\mbox{\boldmath ${\cal E}_{pq}$}
=
\mbox{\boldmath ${\cal A}^{p \dag }_{~r}$}\mbox{\boldmath ${\cal B}_{qr}$}
-
\mbox{\boldmath ${\cal B}^{\mbox{\scriptsize T}}_{qr}$}
\mbox{\boldmath $\bar{\cal A}^{p}_{~r}$}
-
\mbox{\boldmath ${\cal A}^{q \dag }_{~r}$}\mbox{\boldmath ${\cal B}_{pr}$}
+
\mbox{\boldmath ${\cal B}^{\mbox{\scriptsize T}}_{pr}$}
\mbox{\boldmath $\bar{\cal A}^{q}_{~r}$} 
=
\mbox{\boldmath ${\cal A}^{p \dag }_{~\tilde{r}}$}
\mbox{\boldmath ${\cal B}_{q \tilde{r}}$}
-
\mbox{\boldmath ${\cal A}^{q \dag }_{~\tilde{r}}$}
\mbox{\boldmath ${\cal B}_{p \tilde{r}}$} ,
\EA
\right\}
\eeqa
by using the notation
$\!
\mbox{\boldmath ${\cal A}^{p \mbox{\scriptsize T}}_{~r\!+\!N}$} 
\!\!=\!\!
\mbox{\boldmath ${\cal B}^\dag _{pr}$}\!
$
and
$\!
\mbox{\boldmath ${\cal B}^{\mbox{\scriptsize T}}_{pr\!+\!N}$}
\!\!=\!\!
\mbox{\boldmath ${\cal A}^{p \dag }_{~r}$}\!
$
to use a suffix $\tilde{r}$ running from 
0 to $N$ and from $N$ to $2N$.
Then we have the boson images of 
the fermion $SO(2N \!+\!1)$ Lie operators as
\beqa
\left.
\BA{rl}
\mbox{\boldmath $E^\alpha_{~\beta }$}
=
&\!\!\!
\mbox{\boldmath ${\cal E}^\alpha_{~\beta }$}
=
\mbox{\boldmath ${\cal B}^\dag _{\alpha \tilde{r}}$}
\mbox{\boldmath ${\cal B}_{\beta \tilde{r}}$}
-
\mbox{\boldmath ${\cal A}^{\beta \dag }_{~\tilde{r}}$}
\mbox{\boldmath ${\cal A}^\alpha_{~\tilde{r}}$} ,\\
\\
\mbox{\boldmath $E_{\alpha \beta }$}
=
&\!\!\!
\mbox{\boldmath ${\cal E}_{\alpha \beta }$}
=
\mbox{\boldmath ${\cal A}^{\alpha \dag }_{~\tilde{r}}$}
\mbox{\boldmath ${\cal B}_{\beta \tilde{r}}$}
-
\mbox{\boldmath ${\cal A}^{\beta \dag }_{~\tilde{r}}$}
\mbox{\boldmath ${\cal B}_{\alpha \tilde{r}}$} ,\\
\\
\mbox{\boldmath $c_{\alpha }$}
=
&\!\!\!
\mbox{\boldmath ${\cal E}^{\alpha 0}$}
-
\mbox{\boldmath ${\cal E}^\alpha_{~0 }$} 
=
\mbox{\boldmath ${\cal A}^{\alpha \dag }_{~\tilde{r}}$}
(\mbox{\boldmath ${\cal A}^0_{~\tilde{r}}$}
-
\mbox{\boldmath ${\cal B}_{0 \tilde{r}}$})
+
(\mbox{\boldmath ${\cal A}^{0 \dag }_{~\tilde{r}}$}
-
\mbox{\boldmath ${\cal B}^\dag_{0 \tilde{r}}$})
\mbox{\boldmath ${\cal B}_{\alpha \tilde{r}}$} \\
\\[-12pt]
=
&\!\!\!
\sqrt{2}
\left(
\mbox{\boldmath ${\cal A}^{\alpha \dag }_{~\tilde{r}}$}
\mbox{\boldmath ${\cal Y}_{\tilde{r}}$}
+
\mbox{\boldmath ${\cal Y}^{\dag }_{~\tilde{r}}$}
\mbox{\boldmath ${\cal B}_{\alpha \tilde{r}}$}
\right) ,
~~
\mbox{\boldmath ${\cal Y}_{\tilde{r}}$}
\stackrel{\mathrm{def}}{=}
{\displaystyle \frac{1}{\sqrt{2}}}
(\mbox{\boldmath ${\cal A}^0_{~\tilde{r}}$}
-
\mbox{\boldmath ${\cal B}_{0 \tilde{r}}$}) .
\EA
\right\}
\label{bosonimage2}
\eeqa
and 
$
\mbox{\boldmath ${\cal E}^0_{~0}$}
\!=\!
0
$
and
$
\mbox{\boldmath ${\cal E}_{00}$}
\!=\!
0
$.
The last representation for
$\mbox{\boldmath $c_{\alpha }$}$
in
(\ref{bosonimage2})
involves,
in addition to the original
$\mbox{\boldmath $A^{\alpha }_{~\beta }$}$
and 
$\mbox{\boldmath $B _{\alpha \beta }$}$
bosons
and
$\mbox{\boldmath $X _{\alpha }$}$
bosons,
their complex conjugate bosons and the
$\mbox{\boldmath ${\cal Y}_{\tilde{r}}$}$ 
bosons.
The complex conjugate bosons arise from the use of matrix ${\cal G}$
as the variables of representation and the
$\mbox{\boldmath ${\cal Y}_{\tilde{r}}$}$ 
bosons arise from extension of algebra from $SO(2N)$ to $SO(2N+1)$
and embedding of the $SO(2N+1)$ into $SO(2N+2)$.

Using the relations\\[-10pt]
\beq
{\displaystyle \frac{\partial }{\partial {\cal A}^p_{~q}}}
\det {\cal A}
=
({\cal A}^{-1})^{~q}_p
\det {\cal A} ,~~~~
{\displaystyle \frac{\partial }{\partial {\cal A}^p_{~q}}}
({\cal A}^{-1})^{~r}_s
=
-
({\cal A}^{-1})^{~q}_s
({\cal A}^{-1})^{~r}_p ,
\eeq
we get the relations which are valid when operated on functions
on the right coset$\frac{SO(2N+2)}{SU(N+1)}$\\[-14pt]
\beqa
\left.
\BA{ll}
&
{\displaystyle \frac{\partial }{\partial {\cal B}_{pq}}}
=
\sum _{r < p}
({\cal A}^{-1})^{~q}_r 
{\displaystyle \frac{\partial }{\partial {\cal Q}_{pr}}},\\
\\[-12pt]
&
{\displaystyle \frac{\partial }{\partial {\cal A}^p_{~q}}}
=
-
\sum _{s < r < p}
{\cal Q}_{rp}
({\cal A}^{-1})^{~q}_s 
{\displaystyle \frac{\partial }{\partial {\cal Q}_{rs}}}
-
{\displaystyle \frac{i}{2}}
({\cal A}^{-1})^{~q}_p 
{\displaystyle \frac{\partial }{\partial \tau }} ,
\EA
\right\}
\label{differentialformulas}
\eeqa
from which we can derive the expressions
(\ref{SO2Nplus2LieopQ}).

\newpage


\def\thesection{\Alph{section}}
\setcounter{equation}{0}
\renewcommand{\theequation}{\Alph{section}.\arabic{equation}}
\section{Vacuum function for bosons}


~~
We show here that the function 
$\Phi_{00}({\cal G})$ in ${\cal G} \!\in\! SO(2N+2)$
corresponds to the free fermion vacuum function 
in the physical fermion space.
Then the 
$\Phi_{00}({\cal G})$
must satisfy the conditions
\beq
\left(
\mbox{\boldmath ${\cal E}^p_{~q}$} + \frac{1}{2}\delta_{pq}
\right)
\Phi_{00}({\cal G})
=
\mbox{\boldmath ${\cal E}_{pq}$}\Phi_{00}({\cal G})
=
0 ,~~
\Phi_{00}(1_{2N+2})
=
1 .
\label{vacuumcondition}
\eeq
The vacuum function $\Phi_{00}({\cal G})$ which satisfy
(\ref{vacuumcondition})
is given by
$
\Phi_{00}({\cal G}) 
\!=\!
[\det (\bar{\cal A})]^{\frac{1}{2}} ,
$
the proof of which is made easily as follows:
\beqa
\!\!\!\!\!\!\!\!
\BA{ll}
&
\left(
\mbox{\boldmath ${\cal E}^p_{~q}$} 
+ 
{\displaystyle \frac{1}{2}\delta_{pq}}
\right)
[\det (\bar{\cal A})]^{\frac{1}{2}} \\
\\
&
=
{\displaystyle \frac{1}{2}\delta_{pq}}
[\det (\bar{\cal A})]^{\frac{1}{2}}
+
\left(
\bar{\cal B}_{pr}
{\displaystyle \frac{\partial }{\partial \bar{\cal B}_{qr}}}
-
{\cal B}_{qr}
{\displaystyle \frac{\partial }{\partial {\cal B}_{pr}}}
-
\bar{\cal A}^{q}_{~r}
{\displaystyle \frac{\partial }{\partial \bar{\cal A}^{p}_{~r}}}
+
{\cal A}^p_{~r}
{\displaystyle \frac{\partial }{\partial {\cal A}^q_{~r}}}
\right) 
[\det (\bar{\cal A})]^{\frac{1}{2}} \\
\\
&
=
{\displaystyle \frac{1}{2}\delta_{pq}}
[\det (\bar{\cal A})]^{\frac{1}{2}}
\!-\!
\bar{\cal A}^{q}_{~r}
{\displaystyle \frac{\partial }{\partial \bar{\cal A}^{p}_{~r}}}
[\det (\bar{\cal A})]^{\frac{1}{2}} 
\!=\!
{\displaystyle \frac{1}{2}\delta_{pq}}
[\det (\bar{\cal A})]^{\frac{1}{2}}
\!-\!
{\displaystyle 
\frac{1}{2}\frac{1}{[\det (\bar{\cal A})]^{\frac{1}{2}}}}\!
\bar{\cal A}^{q}_{~r}\!
{\displaystyle \frac{\partial }{\partial \bar{\cal A}^{p}_{~r}}}\!
\det (\bar{\cal A}) \\
\\
&
=
{\displaystyle \frac{1}{2}\delta_{pq}}
[\det (\bar{\cal A})]^{\frac{1}{2}}
-
{\displaystyle 
\frac{1}{2}\frac{1}{[\det (\bar{\cal A})]^{\frac{1}{2}}}}
(\bar{\cal A}\bar{\cal A}^{-1})_{qp}
\det (\bar{\cal A}) 
= 0 ,
\EA
\label{vacuum1}
\eeqa
\beqa
\mbox{\boldmath ${\cal E}_{pq}$}
[\det (\bar{\cal A})]^{\frac{1}{2}} 
=
\left(
\bar{\cal A}^{p}_{~r}
{\displaystyle \frac{\partial }{\partial \bar{\cal B}_{qr}}}
-
{\cal B}_{qr}
{\displaystyle \frac{\partial }{\partial {\cal A}^p_{~r}}}
-
\bar{\cal A}^{q}_{~r}
{\displaystyle \frac{\partial }{\partial \bar{\cal B}_{pr}}}
+
{\cal B}_{pr}
{\displaystyle \frac{\partial }{\partial {\cal A}^q_{~r}}}
\right)
[\det (\bar{\cal A})]^{\frac{1}{2}} 
= 0 .
\label{vacuum2}
\eeqa
The vacuum functions $\Phi_{00}(G)$ in $G \!\in\! SO(2N+1)$ and 
$\Phi_{00}(g)$ in $g \!\in\! SO(2N)$
satisfy
\beq
\mbox{\boldmath $c_{\alpha }$}\Phi_{00}(G)
=
\left(
\mbox{\boldmath $E^\alpha_{~\beta }$} + \frac{1}{2}\delta_{\alpha \beta }
\right)
\Phi_{00}(G)
=
\mbox{\boldmath $E_{\alpha \beta }$}\Phi_{00}(G)
=
0 ,~~
\Phi_{00}(1_{2N+1})
=
1 ,
\label{vacuumcondition2}
\eeq
\beq
\left(
\mbox{\boldmath $e^\alpha_{~\beta }$} + \frac{1}{2}\delta_{\alpha \beta }
\right)
\Phi_{00}(g)
=
\mbox{\boldmath $e_{\alpha \beta }$}\Phi_{00}(g)
=
0 ,~~
\Phi_{00}(1_{2N})
=
1 .
\label{vacuumcondition3}
\eeq

By using the $SO(2N+2)$ Lie operators $E^{pq}$, 
the expression
(\ref{SO2Nplus1wf})
for the $SO(2N+1)$ WF $\ket G$ is converted to a form quite
similar to the $SO(2N)$ WF $\ket g$ as
\beq
\ket G
=
\bra 0 U(G) \ket 0
\exp\left(\frac{1}{2} \cdot {\cal Q}_{pq}E^{pq}\right) \ket 0 ,
\label{SO2N+1wf}
\eeq
where we have used the nilpotency relation $(E^{\alpha 0})^2 \!=\! 0$.
Equation
(\ref{SO2N+1wf}) 
leads to the property 
$U(G) \ket 0 \!=\! U({\cal G}) \ket 0$.
On the other hand, from 
(\ref{calApcalBp})
we get
\beq
\det {\cal A}
=
\frac{1+z}{2} \det a ,~~
\det {\cal B}
=
\left\{
\frac{1-z}{2} 
+ 
\frac{1}{2(1+z)}
\left(
x^{\mbox{\scriptsize T}}q^{-1}x - x^\dag x
\right)
\right\}
\det b
=
0 .
\label{detAdeta}
\eeq
Then we obtain the vacuum function $\Phi_{00}({\cal G})$ 
expressed in terms of the K\"{a}hler potential as
\beq
\overline{\bra 0 U({\cal G}) \ket 0}
=
\Phi_{00}({\cal G})
=
\left[\det(\bar{\cal A})\right]^{\frac{1}{2}}
=
e^{-\frac{1}{4}{\cal K}({\cal Q},~{\cal Q}^\dag )}
e^{-i\frac{\tau }{2}} , 
\label{Bogowf2}
\eeq
\beq
\Phi_{00}({\cal G}) 
= 
\Phi_{00}(G)
= 
\sqrt{\frac{1+z}{2}}
\left[\det(\bar{a})\right]^{\frac{1}{2}} 
= 
\sqrt{\frac{1+z}{2}}
e^{-\frac{1}{4}{\cal K}(q,~q^\dag )}
e^{-i\frac{\tau }{2}} .
\label{SO2Nplus1vacuumf2}
\eeq

\newpage

 
\def\thesection{\Alph{section}}
\setcounter{equation}{0}
\renewcommand{\theequation}{\Alph{section}.\arabic{equation}}
\section{\!Differential forms for bosons 
over SO(\!2N+2\!)\!/U(\!N+\!1\!) coset space}


~~~Using the differential 
formulas
(\ref{differentialformulas}),
the fermion $SO(2N+2)$ Lie operators are mapped into
the regular representation consisting of functions 
on the coset space
$\frac{SO(2N+2)}{U(N+1)}$. 
The boson images of the fermion $SO(2N+2)$ Lie operators
\mbox{\boldmath ${\cal E}^p_{~q}$} etc. 
can be 
represented by the closed first order differential forms
over the $\frac{SO(2N+2)}{U(N+1)}$ coset space
in terms of the $\frac{SO(2N+2)}{U(N+1)}$ coset variables ${\cal Q}_{pq}$ 
and 
the phase variable
$
\tau 
\left(
\!=\!
\frac{i}{2} \ln \!
\left[\frac{\det({A}^*)}{\det({A})}
\right] \!
\right)
$  
of the $U(N+1)$,
which is 
identical with the phase variable
$
\tau 
\left(
\!=\!
\frac{i}{2} \ln \!
\left[\frac{\det({a}^*)}{\det({a})}
\right] \!
\right)
$ 
of the $U(N)$
due to the first equation of 
(\ref{detAdeta}), 
as
\beqa
\left.
\BA{ll}
&\mbox{\boldmath ${\cal E}^p_{~q}$} 
=
{\displaystyle
\bar{{\cal Q}}_{pr}\frac{\partial }{\partial \bar{{\cal Q}}_{qr}}
-{\cal Q}_{qr}\frac{\partial }{\partial {\cal Q}_{pr}}
-i\delta_{pq}\frac{\partial }{\partial \tau }
} ,\\
\\
&\mbox{\boldmath ${\cal E}_{pq}$} 
=
{\displaystyle
{\cal Q}_{pr}{\cal Q}_{sq}\frac{\partial }{\partial {\cal Q}_{rs}}
-\frac{\partial }{\partial \bar{{\cal Q}}_{pq}}
-i{\cal Q}_{pq}\frac{\partial }{\partial \tau }
} ,
\EA
\right\}
\label{SO2Nplus2LieopQ}
\eeqa
which are derived in a way quite analogous to 
the $SO(2N)$ case of the fermion Lie operators.
The images of the fermion $SO(2N+1)$ Lie operators 
are represented with the aid of those of the $SO(2N+2)$ operators.
From 
(\ref{SO2Nplus2LieopQ}),
we can get the representations of the $SO(2N+1)$ Lie operators 
in terms of
the variables $q_{\alpha \beta }$ and $r_\alpha$
\cite{Fuk.77}:
\beqa
\!\!\!\!\!\!\!\!
\left.
\BA{ll}
&\E^\alpha_{~\beta }
=
\mbox{\boldmath ${\cal E}^\alpha_{~\beta }$}
=
{\displaystyle
\mbox{\boldmath $e^\alpha_{~\beta }$}
+
\bar{r}_\alpha\frac{\partial }{\partial \bar{r}_\beta }
-
r_\beta\frac{\partial }{\partial r_\alpha } ,~
\mbox{\boldmath $e^\alpha_{~\beta }$}
=
\bar{q}_{\alpha \gamma }\frac{\partial }
{\partial \bar{q}_{\beta \gamma }}
-
q_{\beta \gamma }\frac{\partial }{\partial q_{\alpha \gamma }}
-
i\delta_{\alpha \beta }\frac{\partial }{\partial \tau }
}~,\\
\\
&\E_{\alpha \beta }
=
\mbox{\boldmath ${\cal E}_{\alpha \beta }$}
=
{\displaystyle
\mbox{\boldmath $e_{\alpha \beta }$}
+
(r_\alpha q_{\beta\xi}-r_\beta q_{\alpha \xi })
\frac{\partial }{\partial r_\xi } ,~
\mbox{\boldmath $e_{\alpha \beta }$}
=
q_{\alpha \gamma }q_{\delta \beta }\frac{\partial }
{\partial q_{\gamma \delta }}
-
\frac{\partial }{\partial \bar{q}_{\alpha \beta }}
-
iq_{\alpha \beta }\frac{\partial }{\partial \tau }
} ,
\EA \!\!
\right\}
\label{SO2Nplus1Lieopa}
\eeqa
\beqa
\mbox{\boldmath $c_{\alpha }$}
\!=\!
\mbox{\boldmath ${\cal E}_{0 \alpha }$}
-
\mbox{\boldmath ${\cal E}^0_{~\alpha }$}
\!=\!
{\displaystyle
\frac{\partial }{\partial \bar{r}_\alpha }
+
\bar{r}_\xi \frac{\partial }{\partial \bar{q}_{\alpha \xi }}
+
(r_{\alpha }r_\xi -q_{\alpha \xi })\frac{\partial }{\partial r_\xi }
-
q_{\alpha\xi }r_\eta\frac{\partial }{\partial q_{\xi \eta }}
+
ir_{\alpha }\frac{\partial }{\partial \tau }
} ,~
\mbox{\boldmath $c^\dagger_{\alpha }$}
\!=\!
-
\mbox{\boldmath $\bar{c}_{\alpha }$} .
\label{SO2Nplus1Lieopb}
\eeqa

The vacuum function $\Phi_{00}(G)$ in $G \!\in\! SO(2N+1)$
is given in
(\ref{vacuumcondition2}). 
Using the relations
\beq
\mbox{\boldmath $c_{\alpha }$} \Phi_{00}(G)
=
0,~~
\mbox{\boldmath $c^{\dag }_{\alpha }$} \Phi_{00}(G)
=
\bar{r}_\alpha \Phi_{00}(G) ,
\label{fermiannhicreaction0}
\eeq
and the property 
$U({\cal G}) \ket 0 \!=\! U(G) \ket 0$
((\ref{SO2N+1wf}) and (\ref{SO2Nplus1vacuumf2})),
we have a
relation
\beq
\mbox{\boldmath $c_{\alpha }$}U({\cal G}) \ket 0
=
-
q_{\alpha \xi } r_{\eta }c^{\dag }_{\xi }c^{\dag }_{\eta }
\cdot U(G) \ket 0
-
\left(
r_{\alpha }
+
q_{\alpha \xi }c^{\dag }_{\xi }
\right) \!
\cdot 
\bar{\Phi }_{00}(G)
\exp
\left(
\frac{1}{2} \cdot q_{\gamma \delta }E^{\gamma \delta }
\right)
\ket 0 ,
\label{fermiactionrel}
\eeq
from which  
we easily get the exact identities
\beq
\mbox{\boldmath $c_{\alpha }$}U({\cal G}) \ket 0
\!=\!
\left(
-r_{\alpha }
+
r_{\alpha }r_{\xi }c^{\dag }_{\xi }-q_{\alpha \xi }c^{\dag }_{\xi }
\right) \!
\cdot 
U(G) \ket 0,~~
\mbox{\boldmath $c^{\dag }_{\alpha }$}U({\cal G}) \ket 0
\!=\!
-c^{\dag }_{\alpha }\cdot U(G) \ket 0 .
\label{fermiaction}
\eeq  
Successively using these identities, on the $U({\cal G}) \ket 0$, 
the operators 
\mbox{\boldmath $c_{\alpha}$} and \mbox{\boldmath $c^{\dag }_{\alpha }$} 
are shown to satisfy exactly the anti-commutation relations of the fermion 
annihilation-creation operators
\cite{Ni.98}:
\beq
\left(
\mbox{\boldmath $c^{\dag }_{\alpha }$}\mbox{\boldmath $c_{\beta }$}
+
\mbox{\boldmath $c_{\beta }$}\mbox{\boldmath $c^{\dag }_{\alpha }$}
\right) 
U({\cal G}) \ket 0
=
\delta_{\alpha \beta }\cdot U({\cal G}) \ket 0 ,
\label{anti-commurel1}
\eeq

\beq
\left(
\mbox{\boldmath $c_{\alpha }$}\mbox{\boldmath $c_{\beta }$}
+
\mbox{\boldmath $c_{\beta }$} \mbox{\boldmath $c_{\alpha }$}
\right)
U({\cal G}) \ket 0
=
(
\mbox{\boldmath $c^{\dag }_{\alpha }$}\mbox{\boldmath $c^{\dag }_{\beta }$}
+
\mbox{\boldmath $c^{\dag }_{\beta }$}\mbox{\boldmath $c^{\dag }_{\alpha }$}
)
U({\cal G}) \ket 0
= 0.
\label{anti-commurel2}
\eeq

\newpage


\def\thesection{\Alph{section}}
\setcounter{equation}{0}
\renewcommand{\theequation}{\Alph{section}.\arabic{equation}}
\section{Euler-Lagrange equation of motion for Hamiltonian}


\def\erwt#1{{<\!\!#1\!\!>_G}}
\def\erwt#1{{<\!\!#1\!\!>_{G(t)}}}
\def\erwtG#1{{<\!\!#1\!\!>_G}}
~~~~The expectation values of the fermion Lie operators 
by the $\ket G$ 
are calculated easily as
\beq
\erwt {E^{\alpha }_{~\beta }+\frac{1}{2} 
\delta _{\alpha \beta }}
=
\bar{\cal R}_{\alpha \beta },~~~
\erwt {E_{\alpha \beta }}
=
-{\cal K}_{\alpha \beta },
~~~
\erwt {E^{\alpha \beta }}
=
\bar{{\cal K}}_{\alpha \beta },
\eeq
\vspace{-0.5cm}
\beq
\erwt {c_{\alpha }}
=
{\cal K}_{\alpha 0}- \bar{{\cal R}}_{\alpha 0},~~~
\erwt {c^{\dag }_{\alpha }}
=
\bar{{\cal K}}_{\alpha 0}-{\cal R}_{\alpha 0},
\eeq
where matrix elements ${\cal R}_{pq}$ and ${\cal K}_{pq}$ are given 
in terms of 
the $\frac{SO(2N+2)}{U(N+1)}$ coset variable 
${\cal Q}_{pq}$ as
\beq
{\cal R}_{pq}
\!=\!
\left[
{\cal Q}(1_{N+1} + {\cal Q}^\dag {\cal Q})^{-1}{\cal Q}^\dag 
\right]_{pq} 
({\cal R}^\dagger
\!=\!
{\cal R}) ,~~
{\cal K}_{pq}
\!=\!
\left[
{\cal Q}(1_{N+1} + {\cal Q}^\dag {\cal Q})^{-1}
\right]_{pq} 
({\cal K}^{\mbox{\scriptsize T}}
\!=\!
-{\cal K}) .
\label{RpqKpq}
\eeq
Using the above expressions, 
the following relations
and derivative formulas
are easily proved:
\beqa
{\cal Q}(1_{N+1} - {\cal R})
=
(1_{N+1} - \bar{{\cal R}}){\cal Q}
=
{\cal K} ,~~
\bar{Q} {\cal K}
=
\bar{{\cal K}} {\cal Q}
=
{\cal R} ,
\label{RpqKpqrel}
\eeqa
\vspace{-0.5cm}
\beqa
\left.
\BA{ll}
&{\displaystyle \frac{\partial {\cal R}_{pq}}{\partial {\cal Q}_{uv}}}
=
- \bar{{\cal K}}_{pu}(1_{N+1} - {\cal R})_{vq} 
+ \bar{{\cal K}}_{pv}(1_{N+1} - {\cal R})_{uq} ,\\
\\[-8pt]
&{\displaystyle \frac{\partial {\cal R}_{pq}}{\partial \bar{{\cal Q}}_{uv}}}
=
-(1_{N+1} - {\cal R})_{pu}{\cal K}_{vq} 
+
(1_{N+1} - {\cal R})_{pv}{\cal K}_{uq},\\
\\[-8pt]
&{\displaystyle \frac{\partial {\cal K}_{pq}}{\partial {\cal Q}_{uv}}}
=
(1_{N+1} - \bar{{\cal R}})_{pu}(1_{N+1} - {\cal R})_{vq} 
- 
(1_{N+1} - \bar{{\cal R}})_{pv}(1_{N+1} - {\cal R})_{uq},\\
\\[-8pt]
&{\displaystyle \frac{\partial {\cal K}_{pq}}{\partial \bar{{\cal Q}}_{uv}}}
=
{\cal K}_{pu}{\cal K}_{vq} - {\cal K}_{pv}{\cal K}_{uq} .
\EA
\right\}
\label{RpqKpqderi}
\eeqa
Using the property 
$U(G) \ket 0 \!=\! U({\cal G}) \ket 0$
($G \!\in\! SO(2N+1),~ 
{\cal G} \!\in\! SO(2N+2)$) and 
(\ref{SO2N+1wf}), 
a classical Lagrangian of a system embedded into the $SO(2N+2)$ group 
is given as
\beqa
\BA{ll}
L({\cal G}(t))
&=
\bra0
U^\dagger({\cal G}(t))
\left\{
i\hbar{\displaystyle \frac{\partial }{\partial t}} - H
\right\}
U({\cal G}(t))
\ket0 \\
\\[-12pt]
&=
{\displaystyle \frac{i\hbar }{2}}\mbox{tr}
\{
(1 + {\cal Q}^\dagger{\cal Q})^{-1}
({\cal Q}^\dagger\dot{\cal Q} - \dot{\cal Q}^\dagger{\cal Q})
\}
-{\displaystyle \frac{\hbar }{2}} \dot \tau -\erwt H~.
\EA
\label{Lagrangian}
\eeqa
The Euler-Lagrange EOM for the $\frac{SO(2N+2)}{U(N+1)}$ coset
variables is calculated to be
\beqa
\!\!\!\!\!\!\!\!
\BA{ll}
&{\displaystyle \frac{d}{dt}}
\left(
{\displaystyle \frac{\partial L}{\partial \dot{\bar{\cal Q}}}}
\right)
-
{\displaystyle \frac{\partial L}{\partial \bar{\cal Q}}}
=
-{\displaystyle \frac{1}{2}} i\hbar
\left[
\dot{\cal Q}(1 + {\cal Q}^\dag {\cal Q})^{-1}
+
(1 + {\cal QQ}^\dag)^{-1}\dot{\cal Q}
\right.\\
\\[-12pt]
&\left.
-
{\cal Q}(1 + {\cal Q}^\dag {\cal Q})^{-1}{\cal Q}^\dag \dot{\cal Q}
(1 + {\cal QQ}^\dag )^{-1}
-
(1 + {\cal QQ}^\dag )^{-1}\dot{\cal Q}{\cal Q}^\dag
(1 + {\cal QQ}^\dag )^{-1}{\cal Q}
\right] \!
+
{\displaystyle \frac{\partial \erwtG {H}} 
{\partial{\bar{\cal Q}}}}
=
0 ,
\EA
\label{Euler-Lagrangeequation}
\eeqa
and its complex conjugate.
The Hamiltonian $H$ consists of one-body and two-body operators.
From 
(\ref{Euler-Lagrangeequation}), 
we obtain 
the Euler-Lagrange EOM for the variable ${\cal Q}$
as  
\beqa
\!\!\!\!\!\!\!\!
\left.
\BA{ll}
&\dot{\cal Q}
=
-{\displaystyle \frac{i}{\hbar }}
\left(1 - \bar{{\cal R}}\right)^{-1}
{\displaystyle 
\frac{
\left(
\partial \erwtG {H} 
+\frac{1}{2} M_\alpha \erwtG {c^\dagger_{\alpha }}
+\frac{1}{2} \bar{M}_\alpha \erwtG {c_{\alpha }}
\right)
}
{\partial \bar{\cal Q}}
}
\left(1 - {\cal R}\right)^{-1},\\
\\[-10pt]
&\langle H\rangle _{G}
\!=\!
h_{\alpha \beta }{\cal R}_{\alpha \beta }
\!+\!
{\displaystyle \frac{1}{2}}
[\alpha \beta|\gamma \delta]
\left(
{\cal R}_{\alpha \beta }{\cal R}_{\gamma \delta }
\!-\!
{\displaystyle \frac{1}{2}}
\bar{{\cal K}}_{\alpha \gamma }{\cal K}_{\delta \beta }
\right) ,~
M_{\alpha }
\!=\!
k_{\alpha \beta } \erwtG {c_{\beta }}
\!+\!
l_{\alpha \beta } \erwtG {c^\dagger_{\beta }} ,
\EA
\right\}
\label{equmotion}
\eeqa
in which a classical Hamiltonian function accompanies additional terms
appearing as a classical part of Lagrange multiplier terms
$\!k_{\alpha \! \beta }\!$ 
and 
$\!l_{\alpha \! \beta }\!$
to select out a spinor sub-space.\hspace{-0.15cm}
Using
(\ref{RpqKpqderi}), 
(\ref{equmotion})
is transformed into an 
EOM
for HB amplitudes
which involve effects of unpaired modes as well as of paired modes.
With the aid of the generalized density matrix
(\ref{densitymat}), 
we thus derive the ETDHB equation in which both modes are treated in a 
unified way
\cite{FYN.77,Fuk.77,Ni.98}.

\newpage


\def\thesection{\Alph{section}}
\setcounter{equation}{0}
\renewcommand{\theequation}{\Alph{section}.\arabic{equation}}
\section{Derivation of (\ref{inverse1plusQQ1f}) and 
(\ref{inverse1plusQQsub23f})}


~~~~Using the representation for ${\cal Q}_f$ given in
(\ref{RRTChif}),
the inverse of the matrix ${\cal X}_f$, ${\cal X}_f ^{-1}$, 
is expressed as
\beq
{\cal X}_f ^{-1}
=
\left[ 
\BA{cc}
\chi_f ^{-1} + r_f r_f ^{\dag } &
- fq \bar{r}_f  \\
\\  
- f r^{\mbox{\scriptsize T}}_f q^{\dag } & 
1 + r_f ^{\dag } r_f
\EA
\right] ,~~
\chi_f ^{-1}
=
1_N + f^2 qq^{\dag } ,
\label{inversematrixX}
\eeq
from which the inverse matrix 
${\mathcal X}_f~
(
=
[1_{N+1} + f^2 {\cal Q}_f{\cal Q}_f ^{\dag }]^{-1}
)$ 
in
(\ref{inverse1plusQQf})
is given as
\beq
{\mathcal X}_f
=
\left[ 
\BA{cc}
{\cal Q}_{f qq^\dag } & {\cal Q}_{f q r} \\
\\ \\ \\
{\cal Q}_{f q r}^{\dag } & {\cal Q}_{f r^\dag r}
\EA 
\right] ,~~
\BA{c}
{\cal Q}_{f qq^\dag }
=
\left[
\chi_f ^{-1} + r_f r_f ^{\dag }
-
{\displaystyle
\frac{f^2}{1 + r_f ^{\dag } r_f}
}
q \bar{r}_f r^{\mbox{\scriptsize T}}_f q^{\dag }
\right]^{-1} , \\
{\cal Q}_{f q r}
=
{\displaystyle
\frac{f}{1 + r_f ^{\dag } r_f}
}
{\cal Q}_{f qq^\dag }^{\dag }
q \bar{r}_f , \\
{\cal Q}_{f r^\dag r}
=
{\displaystyle
\frac{f}{1 + r_f ^{\dag } r_f}
}
\left(
1 + f r^{\mbox{\scriptsize T}}_f q^{\dag } {\cal Q}_{f q r}
\right) .
\EA
\label{1plusQQf}
\eeq
Let us introduce an $N \!\times\! N$ matrix $Y_f$ defined in the form
\beq
Y_f
=
\chi_f r_f r_f ^{\dag }
-
{\displaystyle
\frac{f^2}{1 + r_f ^{\dag } r_f}
}
\chi_f
q \bar{r}_f r^{\mbox{\scriptsize T}}_f q^{\dag } .
\label{definitionYf}
\eeq
The ${\cal Q}_{f qq^\dag }$
in
(\ref{1plusQQf})
is related to $Y_f$ by
\beq
{\cal Q}_{f qq^\dag }
=
(1_N + Y_f )^{-1} \chi_f .
\label{relationQfqqandYf}
\eeq
By repeated uses of both the relation and the identity
\beq
r_f ^{\dag } \chi_f r_f
=
{\displaystyle
\frac{1 - Z^2}{Z^2}
} ,~~
r^{\mbox{\scriptsize T}}_f q^{\dag } \chi_f q \bar{r}_f
=
{\displaystyle \frac{1}{f^2}}
\left(
1 + r_f ^{\dag } r_f 
-
{\displaystyle \frac{1}{Z^2}}
\right) ,~~
r^{\mbox{\scriptsize T}}_f q^{\dag } \chi_f r_f
=
r^\dag_f \chi_f q \bar{r}_f
=
0 ,
\label{relationfandidentityf}
\eeq
we can compute $Y_f ^n~(n \ge 1)$ as 
\beq
Y_f ^n
=
\left(
{\displaystyle
\frac{1 - Z^2}{Z^2}
}
\right)^{n-1}
\chi_f r_f r_f ^{\dag }
-
(-1)^{n-1}
\left(
1 
-
{\displaystyle \frac{1}{1 + r_f ^{\dag } r_f}}
{\displaystyle \frac{1}{Z^2}}
\right)^{n-1}
{\displaystyle
\frac{f^2}{1 + r_f ^{\dag } r_f}
}
\chi_f
q \bar{r}_f r^{\mbox{\scriptsize T}}_f q^{\dag } ,
\label{definitionYfn}
\eeq
Using the formula for an infinite summation,
an inverse matrix
$(1_N + Y_f)^{-1}$
is calculated as
\beqa
\BA{ll}
&(1_N + Y_f)^{-1}
=
1_N
+
\sum_{n=1}^{\infty }(-1)^n Y_f ^n \\
=&\!\!\!\!
1_N
-
\sum_{n=0}^{\infty }
\left(
{\displaystyle
-
\frac{1 - Z^2}{Z^2}
}
\right)^n
\chi_f r_f r_f ^{\dag }
+
\sum_{n=0}^{\infty }
\left(
1 
-
{\displaystyle \frac{1}{1 + r_f ^{\dag } r_f}}
{\displaystyle \frac{1}{Z^2}}
\right)^n
{\displaystyle
\frac{f^2}{1 + r_f ^{\dag } r_f}
}
\chi_f
q \bar{r}_f r^{\mbox{\scriptsize T}}_f q^{\dag } \\
=&\!\!\!\!
1_N
-
Z^2 \chi_f r_f r_f ^{\dag }
+
Z^2 f^2 
\chi_f
q \bar{r}_f r^{\mbox{\scriptsize T}}_f q^{\dag } .
\EA
\label{summationYfn}
\eeqa
The geometric series expansion appearing in 
the right-hand side of the first and second lines of
(\ref{summationYfn})
converge
only if $Z^2$ is sufficiently close to 1.
However, by direct calculation of
$(1_N \!+\! Y_f)(1_N \!+\! Y_f)^{-1}$
using
(\ref{definitionYf}),
the last equation of (\ref{summationYfn})
and 
(\ref{relationfandidentityf}),
it is proved to be $1_N$. 
Then one can see that
the expression obtained for the matrix
$(1_N \!+\! Y_f)^{-1}$
in
(\ref{summationYfn})
is indeed true for all 
$0 \!\leq\! Z^2 \!\leq\! 1$.
Substituting
(\ref{summationYfn})
into
(\ref{relationQfqqandYf}) and (\ref{1plusQQf}),
we can get
(\ref{inverse1plusQQ1f}) and (\ref{inverse1plusQQsub23f}).
If we put $f = 1$,
we also get
(\ref{inverse1plusQQ1}) and (\ref{inverse1plusQQsub23}).

\newpage


\end{document}